\documentclass[aps,prd,twocolumn,showpacs,preprintnumbers]{revtex4}

\usepackage{dcolumn}
\usepackage{bm}
\usepackage{amssymb}
\usepackage{amsmath}
\usepackage[english]{babel}
\usepackage[dvips]{graphicx}
\usepackage{subfigure}
\usepackage{color}
\usepackage{psfrag}

\providecommand{\ppmin}{\phi^+_{\scriptscriptstyle <}}
\providecommand{\ppmag}{\phi^+_{\scriptscriptstyle >}}
\providecommand{\pmmin}{\phi^-_{\scriptscriptstyle <}}
\providecommand{\pmmag}{\phi^-_{\scriptscriptstyle >}}
\providecommand{\ppmmin}{\phi^\pm_{\scriptscriptstyle <}}
\providecommand{\ppmmag}{\phi^\pm_{\scriptscriptstyle >}}
\providecommand{\pdmin}{\phi^\Delta_{\scriptscriptstyle <}}

\providecommand{\pd}{\phi^\Delta}

\providecommand{\varpd}{\varphi^\Delta}
\providecommand{\varpc}{\varphi^C}
\providecommand{\pddot}{\dot\varphi^\Delta}

\newcommand{\mat}[2]{\left[\!\!\begin{array}{#1} #2 \end{array}\!\!\right]}
\newcommand{\dx}{d^4\!x}
\newcommand{\dk}{d\mathbf{k}}
\newcommand{\de}{\partial}
\newcommand{\dd}[1]{\frac{\partial^{#1}}{\partial\delta^{#1}}}

\newcommand{\eps}{\epsilon}
\newcommand{\<}{\langle}
\renewcommand{\>}{\rangle}
\newcommand{\df}{\delta\varphi}
\newcommand{\al}{\alpha}
\newcommand{\B}[1]{\mathbf{B}_{#1}}
\newcommand{\AB}[1]{\mathbf{AB}_{#1}}
\newcommand{\ABA}[1]{\mathbf{AB}_{#1}\mathbf{A}}

\newcommand{\Ainv}{\mathbf{A}^{-1}}
\newcommand{\tr}{\mathrm{Tr}}
\newcommand{\ts}{{t_{60}}}
\newcommand{\tin}{{t_{in}}}
\newcommand{\re}{\mathrm{Re}}
\newcommand{\im}{\mathrm{Im}}

\begin{document}

\title{Influence of Super-Horizon Scales on Cosmological Observables
      \\ Generated during Inflation} 

\author{Sabino Matarrese}
\affiliation{Dipartimento di Fisica `G. Galilei', Universit\`{a} di Padova \\
INFN - Sezione di Padova \\ via F. Marzolo 8, I-35131 Padova, Italy}
\email{matarrese@pd.infn.it}
\author{Marcello A. Musso}
\affiliation{Dipartimento di Fisica Nucleare e Teorica,
Universit\`{a} degli Studi di Pavia \\ INFN - 
Sezione di Pavia \\ via U. Bassi 6, I-27100, Pavia, Italy}
\email{marcello.musso@pv.infn.it}
\author{Antonio Riotto}
\affiliation{INFN - Sezione di Padova \\ via F. 
Marzolo 8, I-35131 Padova, Italy}
\email{antonio.riotto@pd.infn.it}

\date{\today}

\begin{abstract}
\noindent
Using the techniques of out-of-equilibrium field theory, we study the 
influence on the properties of cosmological perturbations generated 
during inflation on observable scales coming from fluctuations corresponding 
today to scales much bigger than the present Hubble radius. 
We write the effective action for the coarse-grained 
inflaton perturbations integrating out the sub-horizon modes, which manifest 
themselves as a colored noise and lead to memory effects.
Using the simple model of a scalar field with cubic self-interactions 
evolving in a fixed de Sitter background, we evaluate the 
two- and three-point correlation function on observable scales.
Our basic procedure shows that perturbations do preserve some memory of the 
super-horizon-scale dynamics, in the form of scale-dependent imprints in the
statistical moments. In particular, we find a {\it blue} tilt of the 
power-spectrum on large scales, in agreement with the recent 
results of the {\it WMAP} collaboration which show a suppression of the 
lower multipoles in the Cosmic Microwave Background anisotropies, 
and a substantial enhancement of the intrinsic non-Gaussianity on 
large scales.  

\end{abstract}


\preprint{DFPD 03/A/42; FNT/T 2003/14}

\maketitle

\numberwithin{equation}{section}

\section{Introduction}

Stochastic inflation provides an efficient approach to study inflationary 
dynamics and has become a very popular way to describe the growth of 
density perturbations on scales larger than the Hubble radius. In the 
first fundamental works \cite{starobinski,Rey,goncharov,Nakao,Nambu,Nambu2,
Linde}, the inflaton field was split into a super-horizon and a 
sub-horizon part directly in the equation of motion. This splitting is 
operated in Fourier space through a window function, that separates 
high from low frequencies. The relevant variable is the long-wavelength 
part, while the sub-horizon modes are collected in an effective noise 
term, playing the role of a classical perturbation to the super-horizon 
dynamics. 

The resulting effective equation of motion is then a Langevin-like 
equation analogous to the one describing {\it Brownian motion}, where the 
deterministic evolution is influenced and modified by the stochasticity 
of the source, whose effects can be taken into account only as a statistical 
average over time. Indeed, in this formalism there is not any knowledge 
of the exact form of the noise, but only of its statistical properties.
 
A more general approach exploits the influence functional method 
\cite{morikawa,Hu2}, and operates the frequency splitting at the action level 
getting rid of the high frequencies via a path-integral over the 
sub-horizon part of the field. The effective action thus obtained contains 
some extra terms that can be interpreted as the coupling of the 
super-horizon field with a classical random noise source, whose 
configurations are statistically weighted by an appropriate functional 
probability distribution, becoming the origin of the stochastic character 
of the Langevin-like equation of motion.

The super-horizon degrees of freedom are then treated as a purely 
classical field, all the quantum fluctuations being collected in the 
classical noise term. This feature is claimed not to be a simple 
computational trick, but an intrinsic characteristic of the system. 
Stochasticity is thus not only a clue to understand the properties of 
inflation and the origin of the large-scale structure in the 
Universe, but also as a way to explain the transition from a quantum to 
a classical world \cite{Polarski}.
From a formal point of view, the quantum decoherence process in 
the stochastic inflation framework has been discussed in various works 
\cite{Habib,decoherence,argentini1,Matacz,Kiefer}, where it was 
pointed out that the classicality of the coarse-grained field (implicitely 
assumed in the first papers) is not necessarily assured, but is subject to 
some restrictive conditions.

Using standard techniques of stochastic processes \cite{fokkerplanck,chandra},
the Langevin equation for the field expectation value leads to 
an evolution equation (the Fokker-Planck equation) for its probability 
distribution function. In the first works, the noise correlation time is 
assumed to be infinitesimally short, and the correlation function for 
different times can therefore be considered as being proportional to Dirac's 
delta function $\delta(t-t')$, which sets its {\it white-noise} properties.  
This assumption allows to apply a well-known formalism for the derivation
of the Fokker-Planck equation and its solution. 
However, the characteristic of the correlation function strongly depend
on the window function, whose choice is not a mere mathematical 
tool, but has several physical effects \cite{argentini}. A white noise 
arises only as a consequence of a sharp momentum cutoff, whereas a smooth 
weighting avoids highly singular noise correlators and produces a 
{\it colored noise}. 

The choice of a colored noise is interesting for at least two reasons. The 
first is that a sharp momentum-space cutoff seems rather unphysical, 
while a smooth weighting of the modes is much more likely. 
Actually, the most natural way to integrate out the small-scale fluctuations 
is to average the field in configuration space, choosing an appropriate 
finite volume window function. 
In most cases, this choice results in a smooth weighting in Fourier space 
(thus producing a colored noise), while the sharp momentum-space 
cutoff corresponds to a rather complicated infinite volume window function 
in configuration space. Moreover, it is possible to single out a wide class of 
window functions for which the shape of the colored noise correlation is 
asymptotically the same \cite{Winitzki}. A second reason may be the fact that a
colored noise could play, during inflation, an important role in producing 
intrinsically non-Gaussian density fluctuations as initial conditions for the 
subsequent evolution of the large-scale structure of the Universe \cite{Hu}. 

In the simplest single-field slow-roll models of inflation, non-Gaussian 
features in scalar perturbations are produced by either the 
self-interaction of the inflaton field \cite{falk}, which are however 
constrained to be very small by the slow-roll conditions, or by the 
back-reaction of field fluctuations on the background 
metric, whose amplitude is also strongly constrained by the slow-roll 
conditions \cite{glmm,gupta,acqua,malda}. 
It has been shown, however, that the most copious source of large-scale 
non-Gaussianity is stored in the post-inflationary second-order 
evolution of perturbations, which sets in a 
universal level of non-Gaussianity for the gauge-invariant gravitational 
potential, which turns out to be of order unity \cite{enhance,allmodels}.
 
In this paper we point out that there is
another source of  intrinsic, and 
generally scale-dependent non-Gaussianity in the fluctuation pattern, 
which originates from the cross-talk between super and sub-horizon scale 
perturbation modes.
On scales much larger than the Hubble radius, non-Gaussian features 
generally arise as a consequence of the non-linear multiplicative form 
of the Langevin equation, when back-reaction effects are accounted for 
\cite{salopek,mol,yv1,yv2,yv3}. 
However, as discussed in Refs. 
\cite{salopek,yv1,yv2,yv3,hodges,condiniz,Kofman}, 
these effects do not directly reflect into the statistical properties 
of cosmological perturbations on sub-horizon scales. 
Indeed, in order to deal with fluctuations relative to our  
patch of the Universe, one cannot simply perform statistical averages over the 
entire ensemble of possible states, rather one should allow for the 
observed smoothness of our Universe on large scales. 
A possible, though approximate, way to take this constraint into 
account is to replace ensemble averages with averages 
over the conditional probability density functional that fluctuations on 
sub-horizon scales assume a certain value, 
{\it given that} the inflaton field is spatially homogeneous 
at $t=\ts$, {\it i.e.} about 60 e-folds before the end of inflation 
(corresponding to a comoving scale slightly larger than the present 
Hubble radius). 
This is equivalent to set, for the probability distribution of the 
fluctuations, the `initial' condition $P(\df,\ts)=\delta(\df-\df_{60})$ 
\cite{salopek,yv1,yv2,yv3,hodges,condiniz,Kofman,glmm,gupta}.
Although this may happen (and in most models it does) well after the 
beginning of the accelerated expansion, if the noise driving the fluctuations 
is white their evolution is Markovian, implying that any notion of the 
previous history is erased. 
The probability distribution then behaves exactly as if inflation had started 
at that time and the level of the inflaton non-Gaussianity remains 
fully negligible. 
On the contrary, a colored noise has a non-vanishing correlation 
time: because of this fact the inflaton keeps memory of what happened before 
the constraint, and its evolution ceases to be a Markov process. 
In this scenario, the probability distribution evolves in a different way, 
and also higher moments become important. 

Since the solution of the Fokker-Planck equation with colored noise 
carries several complications and is still a partly unknown matter, in this 
paper we followed a different approach, trying to perturbatively determine 
the probability distribution for the inflaton field directly solving the 
Langevin equation in a statistical way.

The plan of the paper is as follows. In Section \ref{derivazeqmoto} we 
briefly describe the derivation of the stochastic equation of motion for 
the inflaton field averaged over super-horizon scales, using the influence 
functional method, and evaluate the dependence of the noise on the choice of 
window function. In Section \ref{spettri} we then choose a specific 
Gaussian shaped filter, obtaining the related (colored) noise correlation 
functions, and the variance and power-spectrum of the coarse-grained 
fluctuation field. In Section \ref{interacting}, after introducing a small 
non-linear (cubic) term in the potential, we evaluate the bispectrum and the 
third moment of the field. In Section \ref{distrprob} we investigate 
the memory effects induced by this colored noise and build up a formalism to 
quantitatively determine the relevance of the non-Gaussian features of 
the distribution and their sensitivity to the times before $\ts$. 
Finally, in Section \ref{conclusioni} we draw our conclusions. Some technical 
aspects of our calculation are reported in five Appendices. 


\section{Effective super-horizon action}
\label{derivazeqmoto}

We consider a background de Sitter space-time, whose metric reads 
\begin{equation}
ds^2=dt^2-a^2(t)d\mathbf{x}^2=a^2(\eta)(d\eta^2-d\mathbf{x}^2),
\end{equation}
where $a(t)=e^{Ht}$ is the scale-factor (the Hubble 
parameter $H\equiv\dot a/a$ is constant in time) and $\eta$ is the 
conformal time defined by $d\eta=dt/a$, {\it i.e.} $\eta=-[Ha(t)]^{-1}$.

The action for the inflaton field is the ordinary action in curved 
space-time for a free scalar field with mass $m$
\begin{equation}
  S[\phi]=\int\!\!\dx\sqrt{-g}\frac{1}{2}\left[g^{\mu\nu}\de_\mu\phi\de_\nu
  \phi-m^2\phi^2\right]
\end{equation}
(where greek letters label space-time indices), that with our choice of 
background de Sitter metric becomes 
\begin{equation}
\label{azione}
  S[\phi]=\int\!\!\dx 
a^3\frac{1}{2}\!\left[(\de_t\phi)^2-\frac{(\nabla\phi)^2}
{a^2}-m^2\phi^2\right].
\end{equation}

The equation of motion for such a field is 
\begin{equation}
  \label{nonperteqmoto}
  \ddot\phi+3H\dot\phi-\frac{\nabla^2\phi}{a^2}+m^2\phi=0
\end{equation}
and the standard solution for the inflaton Fourier modes reads
\begin{equation}
  \label{mnormali}
  \phi_\mathbf{k}(x)=\frac{H}{(2\pi)^{3/2}}\frac{\sqrt{\pi}}{2}
  \left(\frac{1}{aH}\right)^{3/2}H_\nu^{(1)}\!\left(\frac{k}{aH}\right)
  e^{i\mathbf{k\cdot x}},
\end{equation}
where $\nu^2=\frac{9}{4}-\frac{m^2}{H^2}$ and $H_\nu^{(1)}(x)$ are  
Hankel functions of the first kind. In the special case of a massless 
field, $\nu=3/2$ and
\begin{equation}
\phi_\mathbf{k}(x)=\frac{H}{(2\pi)^{3/2}}\frac{1}{\sqrt{2k}}\left(\eta-
\frac{i}{k}\right)e^{i(-k\eta+\mathbf{k\cdot x})}.
\end{equation}

Let us now split the field $\phi$ in two components, dividing the 
short-wavelength normal modes (with wavelength $a/k$ smaller then 
the horizon scale $H^{-1}$) from the long-wavelength ones. The 
short-wavelength part is defined as
\begin{equation}
\ppmmag=\int d\mathbf{k}W(|\mathbf{k}|,t)[\phi_\mathbf{k}(x)a_\mathbf{k} 
+ \phi_\mathbf{k}^*(x)a^\dagger_\mathbf{k}],
\end{equation}
where the window function $W(k,t)$ projects out the long-wavelength modes.
The long-wavelength part is then simply $\phi_<=\phi-\phi_>$. Substituting 
this field decomposition into the action (\ref{azione}) we obtain two 
distinct free actions for the two fields plus an interaction term:
\begin{equation}
  S[\phi_<,\phi_>]=S[\phi_<]+S[\phi_>]+S_{int}[\phi_<,\phi_>]
\end{equation}
In order to obtain real {\it in-in} vacuum expectation values for the field 
$\phi$ instead of the usual {\it in-out} transition amplitudes, the standard 
procedure is to work in the {\it Closed Time Path} (CTP) formalism 
\cite{schwinger} of out-of-equilibrium field theory. Indeed, in ordinary 
quantum field theory, one deals with transition amplitudes in particle 
reactions and one may not study the dynamics of the system. This 
is because one needs the temporal evolution with definite initial conditions 
and not simply the transition amplitudes of particle reactions with fixed 
initial and final conditions. While ordinary quantum field theory yields 
quantum averages of operators evaluated with an {\it in}-state and an 
{\it out}-state, the CTP formalism yields quantum averages of operators 
evaluated in the {\it in}-state without specifying the {\it out}-state. 
In the CTP formalism 
\cite{jordan} the time integration is made along a closed path going from 
the initial time to positive infinity and back to the initial time; the 
path-integral on the field configurations is evaluated on this closed 
path, along which they need not assume the same values on the forward 
and backward branches of the time contour. This is equivalent to 
considering two fields $\phi^+$ and $\phi^-$, with the constraint 
$\phi^+(+\infty)=\phi^-(+\infty)$, with the ordinary single time 
integration on the real axis. These two fields have to be set equal to 
each other in the equation of motion.We thus end up with four different 
fields $\ppmag$, $\ppmin$, $\pmmag$, $\pmmin$ for small and large scales, 
and forward and backward times.

The effective equation of motion for the {\it in-in} expectation value of the 
super-horizon fields $\ppmmin\equiv\varphi^\pm$ can be derived by 
integrating the action over the sub-horizon field variable $\phi_>$. The 
super-horizon effective action obtained in this way contains two extra 
terms in addition to the ordinary action (\ref{azione}), describing 
physical effects related to the horizon crossing of the various normal 
modes \cite{morikawa}. One of them contains both dissipation and non-local 
mass renormalization effects, while the other describes the influence on 
the super-horizon modes by the sub-horizon ones, whose quantum fluctuation 
can be treated as a stochastic noise. This second term is purely 
imaginary, and as such it cannot be interpreted as a standard action: 
indeed, it appears as the result of a statistical weighting over the 
configurations of the stochastic noise fields representing sub-horizon 
quantum fluctuations, which couple to $\varphi$ in the effective 
action $S_\mathrm{eff}[\varphi^\pm]$. 

After some manipulations (see Appendix \ref{appA}, for details), introducing 
two real classical fields $\xi_1$ e $\xi_2$, whose configurations are 
statistically weighted by the probability distribution functional
${\cal P}[\xi_1(x),\xi_2(x)]$, it is actually possible to write
\begin{multline}
\label{gamma}
e^{i\Gamma[\varphi^\pm]}\equiv\int\!\mathcal{D}\ppmmag 
e^{i(S[\ppmin,\ppmag]-S[\pmmin,\pmmag])}=\\
\<e^{iS_\mathrm{eff}[\varphi^\pm]}\>_S=\int\!\mathcal{D}\xi_1 
\mathcal{D}\xi_2 {\cal P}[\xi_1,\xi_2]e^{iS_\mathrm{eff}[\varphi^\pm]},
\end{multline}
where the effective action $S_\mathrm{eff}[\varphi^\pm]$ is (introducing 
$\varpd=\varphi^+-\varphi^-$)
\begin{widetext}
\begin{gather}
S_\mathrm{eff}=S[\varphi^+]-S[\varphi^-]+\int\!\dx \,a^3_t 
H\left[\left(\!\!\left(\frac{3}{2}-\nu\right)\!\varpd(x)+
\frac{\dot\varphi^\Delta(x)}{H}\right)\xi_1(x)+\varpd(x)\xi_2(x)\right].
\end{gather}
The quantum noise on small scales acts then as a classical random 
source. Here, $S$ is simply the free action \eqref{azione}, since the extra 
dissipation and mass renormalization terms are small and can be neglected on a 
first approach, as shown in Appendix \ref{dissipaz}. 

The statistical weight of the two random fields $\xi_1$ and $\xi_2$ is 
Gaussian,
\begin{gather}
\label{pesogaussiano}
P[\xi_1,\xi_2]=\exp\left\{-\frac{H^2}{2}\int\!\dx\dx'[\xi_1(x),\xi_2(x)]
\mathbf{A}^{-1}(x,x')
\left[\!\!
\begin{array}{c}
\xi_1(x')\\ \xi_2(x')
\end{array}\!\!
\right]\right\},
\intertext{where $\mathbf{A}^{-1}(x,x')$ is the 
functional inverse of the symmetric (under simultaneous exchange of 
discrete and continuous indices $i,t\rightarrow j,t'$) kernel}
\label{intdk}
\mathbf{A}(x,x')=\frac{H^6}{8\pi}\!\int\!\!\frac{dk}{k}\: \frac{\sin kr}{kr}
k^5(\eta\eta')^{5/2}W'(k\eta)W'(k\eta')
\mathrm{Re}[\mathbf{M}_\nu(k\eta,k\eta')],
\intertext{with}
\label{Mnu}
\mathbf{M}_\nu(k\eta,k\eta')\equiv
\left[\begin{array}{cc} H_\nu^{(1)}(k|\eta|)H_\nu^{(1)*}(k|\eta'|) & 
-k\eta'H_\nu^{(1)}(k|\eta|)H_{\nu-1}^{(1)*}(k|\eta'|)\\
-k\eta H_{\nu-1}^{(1)}(k|\eta|)H_\nu^{(1)*}(k|\eta'|) & 
k^2|\eta||\eta'|H_{\nu-1}^{(1)}(k|\eta|)H_{\nu-1}^{(1)*}(k|\eta'|)\end{array}
\right].
\end{gather}
\end{widetext}

In the procedure we described, quantum fluctuations of the sub-horizon 
modes are collected via the path-integral in a rapidly varying classical 
noise term coupled to the super-horizon part of the scalar field: these 
fluctuations can thus talk to the super-horizon modes and perturb the 
dynamics on scales larger than the Hubble radius during inflation.  

The coupling of the two random noises $\xi_1$ and $\xi_2$ with the scalar 
field is slightly different from that in Ref. \cite{morikawa}; the 
reason is that the choice of a general window function $W(k\eta)$ 
(which is not necessarily able to produce a sharp cut in the frequencies, 
but can have some spread around the horizon scale) can introduce a 
$k$-dependence in the effective field coupling to the noise, thereby 
spoiling the separation between super- and sub-horizon scales. Our different 
choice of basis avoids this problem. Another consequence of this formulation 
is that the correlation functions of the two noise fields are different 
from each other, and cross-correlations appear, in a similar way to Ref. 
\cite{argentini}. 

The effective equation of motion obtained from this action with the usual 
CTP method is then 
\begin{multline}
0=\left.\frac{\delta S_\mathrm{eff}}{\delta\pd}\right\vert_{\pd=0}
=\ddot\varphi+3H\dot\varphi \\
- \frac{\nabla^2\varphi}{a^2}+m^2\varphi-3H\xi_1-\dot\xi_1+H\xi_2,
\end{multline}
that is a Langevin-like stochastic equation, where the dynamics of the 
field $\varphi$ is subjected to random ``kicks'' given by the rapidly 
varying stochastic force $\xi$. We treat the effect of the random force as 
a perturbation of the classical dynamics and split the field $\varphi$ 
into its mean, obeying the classical equation of motion 
(\ref{nonperteqmoto}), plus a fluctuation $\df$ that by definition 
vanishes when averaged over all noise configurations. In the massless 
case, the equation for the fluctuations becomes 
\begin{gather}
\label{eqmoto}
\ddot\df+3H\dot\df-\frac{\nabla^2\df}{a^2}=3H\xi_1+\dot\xi_1 -H\xi_2;
\intertext{if we neglect the exponentially suppressed spatial gradients, the 
second time derivative $\ddot\df$ (assuming the validity of the slow-roll 
conditions) and the $\dot\xi_1$ term, we finally get}
\label{slowroll}
\dot\df=\xi_1-\frac{\xi_2}{3}\equiv\xi.
\end{gather}
Even in this simple form, a deterministic treatment of this equation is 
impossible, since we do not know the exact form of $\xi_1$ and $\xi_2$. 
However, we can study this equation from a stochastic point of view 
\cite{fokkerplanck,chandra}, in order to understand how the statistical 
properties of the Gaussian noise (that are completely characterized by the 
two-point correlation functions $A_{ij}(x,x')$) determine the behaviour 
of $\df[\xi]$, now treated as a stochastic variable itself. In other 
words, our goal will not be the exact determination of the evolution of 
the field $\df$, but of its probability distribution functional.


\section{Spectra and Two-point Correlation Functions}
\label{spettri}

From the equation of motion \eqref{slowroll}, one can 
immediately compute the two-point correlation function for $\df$, which reads
\begin{equation}
\label{2correl}
  \langle\df(x)\df(x')\rangle = \!\int_{t_{60}}^t \kern-3mm d\tilde t 
  \!\!\int_{t_{60}}^{t'}\kern-3mm d\tilde t' 
  \<\xi(\tilde t,\mathbf{x})\xi(\tilde t',\mathbf{x}')\> \;,
\end{equation}
where the correlation function of the noise $\xi$ is given by
\begin{multline}
\label{corr}
  \langle\xi(x)\xi(x')\rangle=\frac{1}{H^2}\bigg[A_{11}(x,x') - 
  \frac{A_{12}(x,x')}{3} \\ -\frac{A_{21}(x,x')}{3}
  +\frac{A_{22}(x,x')}{9}\bigg],
\end{multline}
and the exact form of the matrix elements $A_{ij}$ depends through 
(\ref{intdk}) on the choice of the window function $W(k\eta)$.

It is interesting to note that if we project the modes using the Heaviside 
step-function $W(k\eta)=\vartheta((k/aH)-\varepsilon)$, in such a way that 
$W'(k\eta)=\delta((k/aH)-\varepsilon)=aH\delta(k-\varepsilon aH)$, the 
noise correlator $\<\xi_1\xi_1\>$ gives the standard result obtained in the
first stochastic inflation works 
\cite{starobinski,Rey,goncharov,Nakao,Nambu,Nambu2,Linde}
\begin{equation}
\<\xi_1(x)\xi_1(x')\>=\frac{H^3}{4\pi^2}\frac{\sin \eps a_tHr}{\eps 
a_tHr}(1+\eps^2)\delta(t-t'),
\end{equation}
while the other correlators are of order $\mathcal{O}(\eps^2)$. Therefore, 
in the small $\eps$ limit, we get the standard result (for 
$\mathbf{x}=\mathbf{x'}$ and $t<t'$)
\begin{equation}
\<\df(t)\df(t')\>=\frac{H^3}{4\pi^2}(t-t_{60}).
\label{2pstandard}
\end{equation}

Integrating out long wavelenghts using the step-function gives what in 
stochastic language is called a {\it Markov process}. 
However, this is not the most 
natural choice one can do. Actually, the smoothing of the field is generally 
performed in configuration space through a function $w(r/R)$ that rapidly 
decays for distances much larger than $R$. In momentum space, this operation 
produces a weighting of the modes with the Fourier transform $\tilde w(kR)$ 
that projects out the high frequencies. Since we are interested in the short 
wavelenghts part of the field, and our smoothing scale is the comoving 
Hubble length $|\eta|$, a natural choice of the momentum window function will
then be $W(k\eta)=1-\tilde w(k|\eta|)$. 

In a general case the choice of a more physical way to separate 
the modes gives a colored noise term, which is unfortunately much more 
difficult to treat. Namely, if we smooth the field with a Gaussian filter 
\begin{gather}
w(r/|\eta|)=e^{-\frac{k^2\sigma^2}{2\eta^2}},
\intertext{we get the window function}
\label{finestra}
W(k\eta)=1-e^{-\frac{k^2\eta^2}{2\sigma^2}};
\end{gather}
also in this case we can compute the noise correlation function for $\nu=3/2$, 
and in the limit $r=|\mathbf{x}-\mathbf{x}'|\rightarrow0$ we obtain (Appendix 
\ref{noisecorr}), setting $\tau=t-t'$,

\begin{widetext}
\begin{gather}
\label{serie}
\<\xi_i(t)\xi_j(t')\>_{r=0} = \frac{H^4}{4\pi^2}\frac{1}{\cosh^2 H\tau}
\sum^\infty_{k=0}\mathcal{A}^{(k)}_{ij}(\tau)\frac{(-1)^k(k+1)(k+2)}{(2k-1)!!}
\left(2\sigma^{2}\frac{\sinh^2\frac{H\tau}{2}}{\cosh H\tau}\right)^k,
\intertext{where}
\label{matrice}
\mathcal{A}^{(k)}_{ij}(\tau)=\frac{1}{2}\mat{cc}{\displaystyle
\frac{1-2k}{k+2}+\frac{\sigma^2}{\cosh H\tau} & \displaystyle 
\frac{\sigma^2e^{H\tau}}{\cosh H\tau}\left(1+\frac{2k}{e^{H\tau}-1}\right) 
\vspace{4pt}\\
\displaystyle  \frac{\sigma^2e^{-H\tau}}{\cosh 
H\tau}\left(1+\frac{2k}{e^{-H\tau}-1}\right) & \displaystyle \sigma^4 
\frac{(k+3)}{\cosh^2 H\tau}}.
\end{gather}
\end{widetext}

\begin{figure}[h]
\psfrag{Ht}{\kern -1em \raisebox{-.25cm}{$H(t-t')$}}
\psfrag{A}{$\kern -1.5em \frac{8\pi^2}{H^4}\<\xi\xi\>$}
\includegraphics[width=.39\textwidth,height=.31\textwidth]{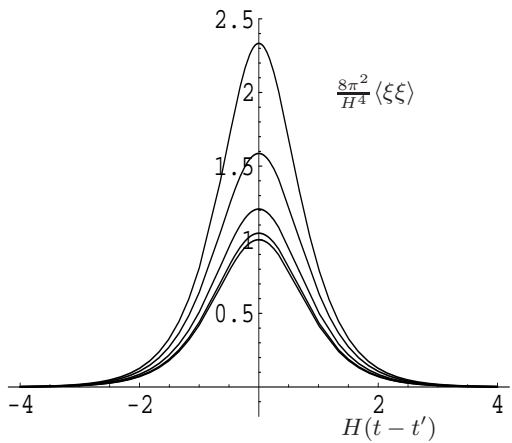}
\label{noisecorrfig}
\caption{Noise correlation functions $\<\xi(t)\xi(t')\>$ for $r=0$ and five 
values of $\sigma$ varying from 1 (top line) to 0 (bottom line). 
The only non-vanishing function for $\sigma=0$ is $\<\xi_1\xi_1\>$}
\end{figure}

For generic values of the parameter $\sigma$, the noise correlation function 
$\<\xi(t)\xi(t')\>$ can have a rather complicated functional form, and is
plotted in Figure \ref{noisecorrfig}. However, in the small $\sigma$ limit, 
the leading term of the series (\ref{serie}) 
is that with $k=0$. Therefore, all the $A_{ij}$ vanish but $A_{11}$, and 
the noise correlation function becomes \cite{Winitzki}
\begin{equation}
\label{limnoisecorr}
\<\xi(t)\xi(t')\> = 
\frac{H^4}{8\pi^2}\frac{1}{\cosh^2\left(H(t-t')\right)}+
\mathcal{O}(\sigma^2),
\end{equation}
which inserted in (\ref{2correl}) and after a double time integration 
gives (to leading order in $\sigma^2$)
\begin{multline}
\label{duepunti}
\<\df(t)\df(t')\> =\frac{H^2}{8\pi^2} \\
\ln\frac{\cosh(H(t-\ts))\cosh(H(t'-\ts))}
{\cosh(H(t-t'))},
\end{multline}
that for $t'\gg t\gg\ts$ approaches the standard result (\ref{2pstandard}). 
Conversely, for $t=t'\rightarrow\ts$ one obtains
\begin{equation}
\label{varianza}
  \<\df(t)^2\>=\frac{H^4}{8\pi^2}\ln\cosh^2(H(t-\ts))
  \sim\frac{H^4}{8\pi^2}(t - \ts)^2 ;
\end{equation}
we thus find that soon after the initial condition is set the variance depends 
quadratically on time, unlike the white-noise case \eqref{2pstandard}, where 
the time dependence is linear. This behaviour, which comes from the fact that 
the Dirac delta function is divergent while a more physical correlation 
function is not, will have an important role in the following analysis.


The power-spectrum can be immediately derived by inserting the explicit 
form (\ref{intdk}) of the $A_{ij}$ into \eqref{2correl}. The two 
integrations over time factor out, and performing the change of variables 
$\tilde t\rightarrow x\equiv-k\tilde\eta$ and $\tilde t'\rightarrow 
y\equiv-k\tilde\eta'$ we get
\begin{equation}
  \<\df(x)\df(x')\>=\int\frac{dk}{k}\frac{\sin kr}{kr}\frac{H^2}{4\pi^2}
  F(k\eta,k\eta'),
\label{2pgauss}
\end{equation}
with
\begin{multline}
\label{spettro}
F(k\eta,k\eta')=\re\!\left[\int_{k|\eta|}^{k|\eta_{60}|} 
\kern -1,5em dx\big(f_1(x)- f_2(x)\big)\right.\\
\times \left.\int_{k|\eta'|}^{k|\eta_{60}|} \kern -1,5em 
dy\big(f_1(y)- f_2(y)\big)^{\!*}\,\right]\!,
\end{multline}
and
\begin{gather}
f_1(x)=-\sqrt{\frac{\pi}{2}}x^{3/2}W'(x)H_{\nu}^{(1)}(x),\\
f_2(x)=-\sqrt{\frac{\pi}{2}} \frac{x^{5/2}}{3} W'(x)H_{\nu-1}^{(1)}(x).
\intertext{For $\nu=3/2$ and using the Gaussian window (\ref{finestra}), 
so that $W'(x)=x\exp[-k^2\eta^2/2\sigma^2]/\sigma^2$, the functions $f_1$ 
and $f_2$ are given by}
f_1(x)=\frac{x(x+i)}{\sigma^2}e^{-\frac{x^2}{2\sigma^2}+ix}, \\
f_2(x)=\frac{ix^3}{3\sigma^2}e^{-\frac{x^2}{2\sigma^2}+ix}.
\intertext{Setting $t=t'$ in \eqref{2pgauss}, we extract the power-spectrum:}
    \mathcal{P}_{\df} = \left(\frac{H}{2\pi}\right)^{\!2} 
  \left|\int_{k|\eta|}^{k|\eta_{60}|}
  \kern -1.1em dx \Big(f_1(x) - f_2(x)\Big)\right|^2 
\end{gather}

For $t\gg\ts$, we can take into account scales that are much larger than the 
horizon at time $t$ (such that $- k\eta \ll 1$) but much smaller at 
$\ts$ ($-k\eta_{60} \gg 1$); for such scales we get 
\begin{equation}
  \mathcal{P}_{\df} \simeq 
  \left(\frac{H}{2\pi}\right)^{\!2} \left|\int_0^\infty
  \kern -1.1em dx f_1(x) - \int_0^\infty \kern -1.1em dx f_2(x)\right|^2
\end{equation}
Studying the behaviour of the integrals of $f_1$ and $f_2$ for large and small 
values of the parameter $\sigma$ we see that for $\sigma\rightarrow0$ we have 
$\int_0^\infty \kern -.3em dx f_1(x)\rightarrow i$ and $\int_0^\infty 
\kern -.3em dx f_2(x)\rightarrow0$, while for 
$\sigma\rightarrow\infty$ both integrals tend to vanish. This means that 
for small $\sigma$ the power-spectrum on very large scales (such that 
$k\ll\sigma a H$) approximates the standard result $H^2/4\pi^2$, while 
fluctuations do not appear at all for very large values of 
the parameter. 
This behaviour is due to the fact that for small $\sigma$ the window 
function tends to unity, and larger and larger scales are included in the 
noise $\xi$, whose correlation function therefore reproduces the ordinary 
fluctuation behaviour (when all scales are taken into account). For 
increasing $\sigma$, instead, since the window function vanishes, the 
noise contains only smaller and smaller wavelengths, and is no more able 
to influence super-horizon scales.


\section{The interacting scalar field}

\label{interacting}

Let us now introduce a small self-interaction term in the potential for 
$\varphi$, in the form of a cubic term 
\begin{equation}
V=\frac{\mu}{3!}\varphi^3,
\end{equation}
where $\frac{\mu}{6H}\ll1$. This toy-model may be useful even to describe
the generation of non-linearities in scalar fields other than the
inflaton field and then trasmitted to the latter, as described in 
Refs. \cite{a1,a2}. We expand to second order in the fluctuations 
the field $\df$ substituting $\df=\df_1+\df_2$ in \eqref{slowroll}. 
Assuming that the small self-interaction term can only produce second-order 
effects (acting therefore only on $\df_2$), we can solve recursively the 
equation of motion for $\delta\varphi_1$ and $\delta\varphi_2$. In the 
slow-roll approximation they become
\begin{equation}
\left\{\begin{aligned}
\label{iteraz1}
& 3H\dot{\df_1} = 3H\xi_1+\dot\xi_1-H\xi_2,\\
& 3H\dot{\df_2} + \frac{\mu}{2}\delta\varphi_1^2= 0.
\end{aligned}\right.
\end{equation}
To first order in the field expansion, the three-point function decomposes 
into the sum of three terms:
\begin{multline}
\label{trepunti}
\<\df_1(x_1)\df_1(x_2)\df_2(x_3)\> 
+ \<\df_1(x_1)\df_2(x_2)\df_1(x_3)\> \\
+ \<\df_2(x_1)\df_1(x_2)\df_1(x_3)\>.
\end{multline}
Since $\df_2$ depends quadratically on $\df_1$, each of these first-order 
terms is actually constituted by the sum of three four-point functions of 
$\df_1$, two connected and one disconnected. However, since in the 
perturbative expansion in $\frac{\mu}{6H}$ of the equation of motion the 
originally vanishing mean value of $\df$ is shifted away from zero by 
$\<\df_2\>$, and the disconnected term is indeed proportional to this 
quantity, we can eliminate it just by requiring that the mean of the 
fluctuation vanishes. For this purpose, we set
\begin{equation}
\label{iteraz3}
  \df_2=-\frac{\mu}{6H}\int_\ts^t dt'[\df_1^2(t',\mathbf{x})-
  \<\df_1^2(t',\mathbf{x})\>];
\end{equation}
the second term, added in order to make the mean value vanish, cancels 
the disconnected contribution, while the connected ones are equal to each 
other. The first term of (\ref{trepunti}), namely, becomes
\begin{multline}
\label{trepunti2}
-\frac{\mu}{3H}\int_\ts^{t_3} 
dt'\<\df_1(t_1,\mathbf{x}_1)\df_1(t',\mathbf{x}_3)\> \\
\times \<\df_1(t_2,\mathbf{x}_2)\df_1(t',\mathbf{x}_3)\>,
\end{multline}
that is (setting $t_1=t_2=t_3\equiv t$ and going to conformal time)
\begin{multline}
\frac{\mu}{3H^2}\int_{\eta_{60}}^{\eta} 
\frac{d\eta'}{\eta'}\frac{H^4}{4(2\pi)^6}\!\int\!\frac{\dk}{k^3} 
e^{i\mathbf{k} \cdot (\mathbf{x}_1-\mathbf{x}_3)}F(k\eta,k\eta')\\
\times \int\!\!\frac{\dk'}{{k'}^3} 
e^{i\mathbf{k}' \cdot (\mathbf{x}_1-\mathbf{x}_3)}F(k'\eta,k'\eta').
\end{multline}

Going to Fourier space and extracting a factor of 
$(2\pi)^3\delta(\mathbf{k}_1+\mathbf{k}_2+\mathbf{k}_3)$, we obtain 
the total bispectrum as the sum of the three 
contributions
\begin{equation}
B(\mathbf{k}_1,\mathbf{k}_2,\mathbf{k}_3)=B_{12}+{\rm permutations}, 
\end{equation}
where 
\begin{multline}
B_{12}\equiv\frac{\mu H^2}{12}\frac{1}{k_1^3k_2^3}\int_{\eta_{60}}^{\eta} 
\frac{d\eta'}{\eta'}\\ 
\times F(k_1\eta,k_1\eta')F(k_2\eta,k_2\eta').
\end{multline}
With good approximation, 
we can write for the Gaussian window case
\begin{multline}
\label{spectgauss}
  F(k\eta,k\eta') \simeq 
  \left(e^{-\frac{k^2\eta^2}{2\sigma^2}}g(k\eta) -
  e^{-\frac{k^2\eta_{60}^2}{2\sigma^2}}g(k\eta_{60})\right)   \\
  \times \left(e^{-\frac{k^2\eta^2}{2\sigma^2}}g(k\eta) -
  e^{-\frac{k^2\eta_{60}^2}{2\sigma^2}}g(k\eta_{60})\right)
  + \mathcal{O}(\sigma^4)
\end{multline}
where
\begin{equation}
  g(x) \equiv 1+\frac{\sigma^2}{3}+\frac{1+\sigma^2}{6}x^2
\end{equation}
and the contributions to the total bispectrum can be now calculated 
analytically. In the limit of super-horizon scales 
($-k_i\eta/\sigma\ll1$, but $-k_i\eta_{60}/\sigma\gg1$) and to order 
$\mathcal{O}(\sigma^4)$ we obtain
\begin{equation}
  B_{12}=\frac{\mu H^2}{24}\frac{1+\sigma^2/3}{k_1^3k_2^3}\bigg[\!
  \ln\frac{(k_1^2+k_2^2)\eta^2}{2\sigma^2}+\gamma - \frac{\sigma^2}{3}\bigg]
\end{equation}

If we take as our window the step-function, so that the derivative 
$W'(x)$ is Dirac's delta function $\delta(x-\varepsilon)$, we get
\begin{multline}
\label{spectdelta}
F(k\eta,k\eta')=\left(1+\frac{\varepsilon^2}{3}+
\frac{\varepsilon^4}{9}\right)\vartheta(\varepsilon-k|\eta|) \\
\vartheta(\varepsilon-k|\eta'|)\vartheta(k|\eta_{60}|-\varepsilon),
\end{multline}
and
\begin{equation}
B_{12}=\frac{\mu H^2}{12}\frac{1}{k_1^3k_2^3}
\left(1+\frac{\varepsilon^2}{3}+\frac{\varepsilon^4}{9}\right)^2
\ln\frac{k_{max}|\eta|}{\varepsilon},
\end{equation}
with $k_{max}|\eta|<\varepsilon<k_{min}|\eta_{60}|$. This result is 
in good agreement with the one in Ref. \cite{falk}, the main difference being 
that the relevant scale is not $k_1+k_2+k_3$ but $k_{max}$, which is due to our
use of the slow-roll approximation, instead of the exact solution of the 
equation of motion. 

For later convenience, let us also calculate here the skewness 
of our coarse grained inflaton field: from (\ref{trepunti}) 
and (\ref{trepunti2}), with $x_1=x_2=x_3$, we have in the white-noise case 
(when $\<\df^2\>\propto t-\ts$) 
\begin{gather}
\label{trepuntistandard}
  \<\df^3(t,\mathbf{x})\>=-\frac{\mu\, H^5}{3\left(2\pi\right)^4}(t-\ts)^3,
\intertext{while with our window function the result, in the small 
$\sigma$ limit, is}
\label{3puntinomemory}
  \<\df^3(t,\mathbf{x})\>=\frac{\mu\,H^2}{4(2\pi)^4}\int_{\eta_{60}}^{\eta}
  \!\!\frac{d\tilde\eta}{\tilde\eta}\!\left[\ln
  \frac{(\eta^2+\eta_{60}^2)(\tilde\eta^2+\eta_{60}^2)}
      {(\eta^2+\tilde\eta^2)2\eta_{60}^2}\right]^2\!;
\intertext{for early times it can be approximated by}
  \<\df^3(t,\mathbf{x})\> \simeq - \frac{\mu 
H^7}{12\left(2\pi\right)^4}(t-\ts)^5,
\end{gather}
while for $t\gg\ts$ we asymptotically recover the ordinary result 
(\ref{trepuntistandard}).


\section{Conditional Probability Distribution and Moments}
\label{distrprob}

In this section we derive an expression for the probability distribution 
of the field $\df$ at time $t$ as a function of the stochastic variable 
$\xi=\xi_1-\xi_2/3$. More precisely, we are interested in the conditional 
probability $P_t(\df|\df_{60})$ that the stochastic variable $\df[\xi](t)$ 
assumes a specific value $\df$ at time $t$ given that it assumed the value 
$\df_{60}$ at the earlier time $t_{60}$, {\it i.e.} 
60 e-folds before inflation ends. In particular, we will consider 
$\df_{60}=0$, as argued in 
\cite{hodges,condiniz,Kofman}. 
This is the crucial quantity that distinguishes Markovian processes from 
non-Markovian ones. A stochastic process is said to be Markovian if its 
conditional probability depends only on the value of the variable at the 
time when we put the constraint, while in non-Markovian cases a memory of 
what happened at earlier times is kept. 

It is a well known result of the theory of stochasticity 
\cite{fokkerplanck,chandra} that Langevin equations with white noise 
describe Markovian processes, while colored noise is a typical cause for 
Markovianity to break down. Thus, in our case the fluctuations of the 
inflaton field on large scale behave like a Markovian process if we choose 
$W(k\eta)$ to be a step-function, that generates white noise, but become 
non-Markovian whenever a smoother separator is applied, as for example our 
Gaussian window function.

Our goal is to estimate the amount of non-Gaussianity produced in this 
latter case: in the standard scenario, non-Gaussian features are generally 
small, since they do not have enough time to develop after $t_{60}$, 
when the condition $\df_{60}=0$ applies. However, if 
the constraint does not erase the memory of earlier times, non-Gaussianity 
could be significantly larger. 

According to Bayes theorem, the conditional probability is obtained from 
the joint probability  $P_t(\df,\df_{60})$ that the random process 
$\df[\xi]$ assumes the values $\df_{60}$ at time $t_{60}$ and $\df$ at 
time $t$, normalised with the probability $P(\df_{60})$ that 
$\df[\xi](t_{60})=\df_{60}$.
A simple way to evaluate the joint probability (see {\it e.g.} 
Ref. \cite{sabino}) is to average over all $\xi$ configurations the product 
of two delta functions centered on these values:
\begin{widetext}
\begin{align}
  P_t\big(\df,\df_{60}\big) &= \mathcal{N}\!\!\int \!\!\mathcal{D}\xi\:
  \delta\Big(\df[\xi](t)-\df\Big)\:\delta\Big(\df[\xi](t_{60})-\df_{60}\Big)
    \:e^{-\frac{1}{2}\xi^T\mathbf{A}^{-1}\xi} \nonumber\\
  &= \int\!\!\frac{d\al_1}{(2\pi)}\frac{d\al_2}{(2\pi)} 
  e^{-i(\al_1\df+\al_2\df_{60})} \mathcal{N}\!\!\int 
  \!\!\mathcal{D}\xi\:e^{-\frac{1}{2}\xi^T\mathbf{A}^{-1}\xi+
   i\al_1\df[\xi](t)+i\al_2\df[\xi](t_{60})}, 
  \label{pcongiunta}
\end{align}
\end{widetext}
where the shorthand notation $\xi^T\mathbf{A}^{-1}\xi$ stands for the double 
integration in the exponent of the Gaussian weight (\ref{pesogaussiano}). 
The probability distribution $P(\df_{60})$ for $\df[\xi]$ at $t_{60}$ can 
be obtained in the same way by averaging just one delta function:
\begin{align}
\label{p60}
  P(\df_{60}) 
  &= \mathcal{N}\!\!\int\!\!\mathcal{D}\xi\:\delta\Big(\df[\xi](t_{60}) 
  - \df_{60}\Big)\:e^{-\frac{1}{2}\xi^T\mathbf{A}^{-1}\xi} \nonumber\\
  &= \int\!\frac{d\al_2}{2\pi} e^{-i\al_2\df_{60}} \mathcal{N}\!\!\int\!\!
  \mathcal{D}\xi\:e^{-\frac{1}{2}\xi^T\mathbf{A}^{-1}\xi+i\al_2\df[\xi]
  (t_{60})}.
\end{align}

The equation of motion for $\df[\xi]$ is a first order differential 
equation, and therefore for a given choice of the function $\xi(t)$ the 
solution depends on the initial condition $\df_{in}$ at $t=t_{in}$
defined to be time at beginning of inflation. 

Since $\df$ is a fluctuation ({\it i.e.} with vanishing mean value), 
we require that 
the solution vanishes at every time when averaged over all $\xi$'s: the 
initial condition must then be $\df_{in}=0$ for continuity.

If we could solve the equation of motion, we could then calculate the 
value of the fluctuation at $t_{60}$. However, because of the presence of 
the delta function, this value is constrained and the path-integral runs 
only over the noise configurations that satisfy the condition on 
$\df[\xi](t_{60})$. Moreover, the value $\df_{60}$ assumed at $t_{60}$ 
constitutes a new initial condition for the equation of motion at later 
times, whose solution thus depends only on the noise configuration after 
$t_{60}$.

Writing the equation of motion in integral form and then solving 
perturbatively by iteration, to first order in $\frac{\mu}{6H}$ we get
\begin{align}
  \label{eqpert1} 
  \df[\xi](t_{60}) &= \int_{t_{in}}^{t_{60}}\!\!\!dt'\xi(t') - 
  \frac{\mu}{6H}\int_{t_{in}}^{t_{60}}\!\!\!dt'\df^2[\xi](t')\\
  &\simeq \int_{t_{in}}^{t_{60}}\!\!\!dt'\xi(t') - \frac{\mu}{6H}
  \int_{t_{in}}^{t_{60}}\!\!\!dt'\bigg(\int_{t_{in}}^{t'}\!\!\!dt''
    \xi(t'')\bigg)^2 \nonumber
\end{align}
that depends on the noise term configuration only up to $t_{60}$, and for 
later times
\begin{align}
  \df[\xi](t)&= \df_{60} + \int_{t_{60}}^{t}\!\!\!dt'\xi(t') - 
  \frac{\mu}{6H}\int_{t_{60}}^{t}\!\!\!dt'\df^2[\xi](t') \nonumber \\
  &\simeq \df_{60} - \frac{\mu}{6H}\df_{60}^2(t-t_{60}) + 
  \int_{t_{60}}^{t}\!\!\!dt'\xi(t') \nonumber \\
  &\qquad - \frac{\mu}{6H}\int_{t_{60}}^{t}\!\!\!
  dt'\bigg(\int_{t_{60}}^{t'}\!\!\!dt''\xi(t'')\bigg)^2,
  \label{eqpert2}
\end{align}
where in the last equality we did not include the first order correction 
to the linear term in $\xi$ because its effect would just be a sub-leading 
contribution to the two-point correlation function.

As we already pointed out, the solution after $t_{60}$ involves only 
integrals over later times: the only possibility for the fluctuation 
field to keep memory of earlier times is then that this configuration 
itself is influenced by the configurations before the constraint, 
which is exactly what happens in the case of colored noise.

We see that in these equations there are both linear and quadratic terms 
in $\xi$: the quadratic terms can be added to $\xi^T\mathbf{A}^{-1}\xi$ of 
Eq. (\ref{pcongiunta}) in such a way as to obtain a modified integration 
kernel and perform the Gaussian integration over the $\xi$'s. 
Skipping all the technicalities (see Appendix \ref{probabilita}), after 
inverting perturbatively the new integration kernel we obtain for the 
conditional probability the following result:
\begin{multline}
\label{pcondiz}
P_t(\df|\df_{60})=\\
\frac{\displaystyle \int\!\!d\al_1d\al_2\: 
e^{-i\df_i\al_i}e^{-\frac{1}{2}C_{ij}\al_i\al_j +i 
\frac{\mu}{6H}D_{ijk}\al_i\al_j\al_k}}{\displaystyle 2\pi \int\!\!d\al_2 
\:e^{-i\df_2\al_2}e^{-\frac{1}{2}C_{22}\al_2^2 
+ i\frac{\mu}{6H}D_{222}\al_2^3}},
\end{multline}
where we adopted a convenient short-hand tensor notation, setting
\begin{gather}
  \begin{gathered}
    \df_1=\df-\df_{60}+\frac{\mu}{6H}\df_{60}^2(t-\ts)\;, \\ \df_2=\df_{60}\;,
  \end{gathered} \\ 
  C_{ij}=\int_{I_i}\!\!dt'\!\int_{I_j}\!\!dt''\<\xi(t')\xi(t'')\> \;,\\
  D_{ijk}=\int_{I_j}\!\!\!d\tilde t 
  \int_{I_i}\!\!\!dt'\!\!\int_{\tilde I_j}\!\!\!dt''
  \<\xi(t')\xi(t'')\>
  \int_{I_k}\!\!\!d\tau'\!\!\int_{\tilde I_j}\!\!\!d\tau''
  \<\xi(\tau')\xi(\tau'')\>
\end{gather}
and $I_1=[t_{60},t]$, $I_2=[t_{in},t_{60}]$, $\tilde I_1=[t_{60},\tilde 
t]$, $\tilde I_2=[t_{in},\tilde t]$ for the integration supports.

For white-noise processes, the only terms that survive are $C_{11}$, 
$C_{22}$, $D_{111}$ and $D_{222}$, while those with mixed indices vanish. 
For instance, we have
\allowdisplaybreaks{
\begin{gather}
  C_{12}\propto\int_{t_{60}}^t\!\!\!dt'\!\int_{t_{in}}^{t_{60}}\!\!\!\!dt''
  \delta(t'-t'')=0, \\
  D_{121}\propto \int_{t_{in}}^{t_{60}}\!\!\!d\tilde 
  t\bigg(\int_{t_{60}}^t\!\!\!dt'\!\int_{t_{in}}^{\tilde 
  t}\!\!\!\!dt''\delta(t'-t'')\bigg)^2=0
\end{gather}}
since the integration supports of the two time variables are disjoint, and 
the same happens for all other mixed terms. In this case, thus, the two 
integrations over $\al_1$ and $\al_2$ in the numerator factor out, and the 
second one simplifies with the one in the denominator. We then get
\begin{align}
  P_t(\df|\df_{60}) &= \frac{1}{2\pi}\!\int\!\!d\al_1 
  \:e^{-i\df_1\al_1}e^{-\frac{1}{2}C_{11}\al_1^2 
  +i\frac{\mu}{6H}D_{111}\al_1^3} \nonumber \\
  &= \frac{\exp\!\left[\frac{\mu}{6H}D_{111}\frac{\partial^3}{\partial\df^3}
  \right]\exp\!\left[-\frac{\df^2_1}{2C_{11}}\right]}{\sqrt{2\pi C_{11}}}
\end{align}
{\it i.e.} a probability distribution function for a Markovian process, 
that does not keep any memory of what happened before the constraint 
(indeed, this function 
contains only integrations over times after $t_{60}$, since no index 2 
appears). To zeroth order in $\frac{\mu}{6H}$, this distribution is 
Gaussian with mean $\df_{60}$ and variance $C_{11}$, which is exactly the 
standard case of equation (\ref{2correl}).

In a case with colored noise, the correlation function is no more a delta 
function but has a finite width, and the coefficients with mixed indices 
no longer vanish even if the two time integration supports are disjoint. 
In the general non-Markovian case no factorization is possible, and after 
integrating over $\al_1$ and $\al_2$ we can write
\begin{widetext}
\begin{gather}
  P_t(\df|\df_{60})=\frac{1}{\sqrt{2\pi C_{1\!1}(1-y)}}\frac{\exp\!\left[
\frac{\mu}{6H}D_{ijk}\frac{\partial^3}{\partial \df_i \partial \df_j\partial 
\df_k}\right]\exp\!\left[-\frac{1}{2}[C^{-1}]_{ij}\df_i\df_j\right]}
 {\exp\!\left[\frac{\mu}{6H}D_{222}\frac{\partial^3}{\partial 
 \df_2^3}\right]\exp\!\left[-\frac{1}{2}\frac{\df_2^2}{C_{22}}\right]},
\intertext{where C is the matrix with elements $C_{ij}$ and}
  [C^{-1}]_{ij}=\frac{1}{1-y}\mat{rl}{\frac{1}{C_{11}}&\frac{-y}{C_{12}}\\
  \frac{-y}{C_{12}}&\frac{1}{C_{22}}} 
  \quad,\quad y=\frac{C_{12}^2}{C_{11}C_{22}}\;.
\end{gather}
It is immediate to check that from this general formula we can get as a 
particular case the Markovian formula by simply setting 
$C_{12}=0$ ({\it i.e.} $y=0$) and $D_{ijk}=0$ for $\{ijk\}\neq\{111\},\{222\}$.

Expanding the derivation operators to first order in $\frac{\mu}{6H}$, after 
some algebra we get 
\begin{multline}
  P_t(\df|\df_{60}) = 
  \frac{\exp\left[-\frac{1}{2}\frac{\left(
  \df-\df_{60}\big(1+\frac{C_{12}}{C_{22}} - 
  \frac{\mu}{6H}\df_{60}(t-\ts)\big)\right)^2}{C_{11}(1-y)}\right]}
  {\sqrt{2\pi C_{11}(1-y)}}\\
  \times\left(1+\frac{\mu}{6H}\left(\frac{D_{ijk}\frac{\partial^3}
{\partial \df_i \partial 
  \df_j\partial 
  \df_k}e^{-\frac{1}{2}[C^{-1}]_{ij}\df_i\df_j}}{e^{-\frac{1}{2}[C^{-1}]_{ij}
  \df_i\df_j}}-\frac{D_{222}\frac{\partial^3}{\partial 
  \df_2^3}e^{-\frac{\df_2^2}{2C_{22}}}}{e^{-\frac{\df_2^2}
{2C_{22}}}}\right)\right).
\end{multline}
\end{widetext}

To zeroth order in $\frac{\mu}{6H}$ we then have again a Gaussian distribution,
but now the mean is
\begin{gather}
\<\df\>|_{NM} = 
\df_{60}\bigg(1+\frac{C_{12}}{C_{22}}\bigg),
\intertext{that is the sum of the constraint value plus an extra term due to 
non-Markovian memory effects, while
the variance becomes} 
  \<\df^2\>|_{NM}=C_{11}(1-y).
\end{gather}
In order to calculate the skewness, we have to take into account all the first 
order contributions. The full calculation gives
\begin{multline}
  \label{trepuntimem}
  \<\df^3\>|_{NM} = 
  \left.\int_{-\infty}^{+\infty}\!\!d\df\:\df^3 P_t(\df|\df_{60})
  \right|_{\df_{60}=0}\\
  = -\frac{\mu}{H}D_{111} (1+K),
\end{multline}
where $K$ is given by
\begin{multline}
K = -\frac{C_{12}}{C_{22}}\frac{2D_{112}+D_{121}}{D_{111}} - 
\frac{C_{11}}{C_{22}}\frac{1-3y}{2}\frac{2D_{122}+D_{212}}{D_{111}} \\
  + \frac{C_{11}C_{12}}{(C_{22})^2}\frac{3-5y}{2}\frac{D_{222}}{D_{111}};
\end{multline}
$y$ and $K$ are then the terms monitoring the importance of memory effects 
induced by the colored noise in the conditional probability
distribution variance and skewness 
respectively. Whenever these coefficients are small (as for 
white-noise cases, when they identically vanish since $C_{12}=0$ and $D_{112}=
D_{121}=D_{122}=D_{212}=0$) we get for the variance $\<\df^2\>|_M=C_{11}$ the 
ordinary result \eqref{2correl}, while Eq. \eqref{trepuntimem} reduces to
\begin{equation}
  \<\df^3\>|_M = -\frac{\mu}{H}\!\int_\ts^t \!\!\!\!d\tilde t
  \bigg(\int_\ts^t\!\!\!\!dt'\!\!\int_\ts^{\tilde t}
  \!\!\!\!dt''\<\xi(t')\xi(t'')\>\!\bigg)^{\!2} \!,
\end{equation}
which is exactly what we obtain from Eq. (\ref{trepunti})-(\ref{trepunti2}) 
with $x_1=x_2=x_3$. If instead $y$ and $K$ are significantly different from 
zero the effect is big, and the procedure of neglecting times
before $\ts$, as we did to derive for example Eqs. \eqref{varianza} and
\eqref{3puntinomemory}, is not correct any more.

We now want to apply the formalism developed so far to the case of 
the Gaussian window (\ref{finestra}). It is a generic feature of 
non-Markovian systems that, since memory effects appear through secular terms
which depend on the whole time interval before the constraint, they increase 
as this interval increases. Conversely, much after the constraint these effects
tend to be erased as a consequence of stochasticity. The results derived in 
Section \ref{spettri} should thus still hold for $t-t_{60}\gg H^{-1}$, while 
we expect the difference from the Markovian case to be maximal for 
$t-t_{60}\ll H^{-1}$ and $t_{60}-t_{in}\gg H^{-1}$.

In this limit, we can compare the white-noise correlation functions 
(\ref{2pstandard}) and (\ref{trepuntistandard}) with the ones just 
derived. For the two-point function, while in the Markovian case we had 
$\<\df^2\>\propto H(t-\ts)$, now, since the noise correlation never 
diverges and $C_{11}$ contains a double integration over time, for 
$t\rightarrow\ts$ we get $\<\df^2\>\propto H^2(t-\ts)^2$, and the ratio of 
the two vanishes.
Precisely, for a value of the parameter $\sigma$ not too close to 1, at 
the $C_{ij}$ read 
\allowdisplaybreaks{
\begin{equation}
  C_{11} \simeq 
\left(\frac{H}{2\pi}\right)^2\ln\frac{\eta^2+\eta_{60}^2}{2\eta\eta_{60}} 
\sim\frac{H^4}{4\pi^2}\frac{(t-\ts)^2}{2},
\end{equation}
\begin{multline}
  C_{12} = C_{21} \simeq 
  \left(\frac{H}{2\pi}\right)^2\frac{1}{2}\ln\frac{2\eta_{60}^2
  (\eta^2+\eta_{in}^2)}{(\eta^2+\eta_{60}^2)(\eta_{60}^2+\eta_{in}^2)}\\
  \sim\frac{H^3}{4\pi^2}\frac{(t-\ts)}{2},
\end{multline}
\begin{multline}
  C_{22} \simeq 
  \left(\frac{H}{2\pi}\right)^2\ln\frac{\eta_{60}^2+\eta_{in}^2}
  {2\eta_{60}\eta_{in}}\\
  \sim\frac{H^2}{4\pi^2}[H(\ts-\tin)-\ln 2].
\end{multline}

\begin{figure*}
\centering
\psfrag{Ht}{$\Delta N$}
  \subfigure{\psfrag{K}{\kern -3em\raisebox{.1cm}{ 
       $(4\pi^2/H^2)\,\<\df^2\>$}}
   \includegraphics[width=.4\textwidth,height=.3\textwidth]{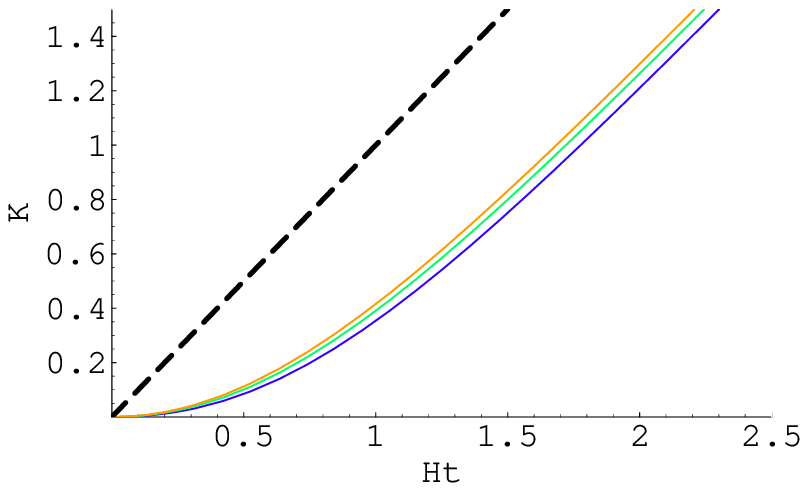}}
  \hspace{.05\textwidth}
  \subfigure{\psfrag{K}{\kern -3em\raisebox{.1cm}{
       $\<\df^2\>|_{NM}\Big/\<\df^2\>|_M$}}
   \includegraphics[width=.4\textwidth,height=.3\textwidth]{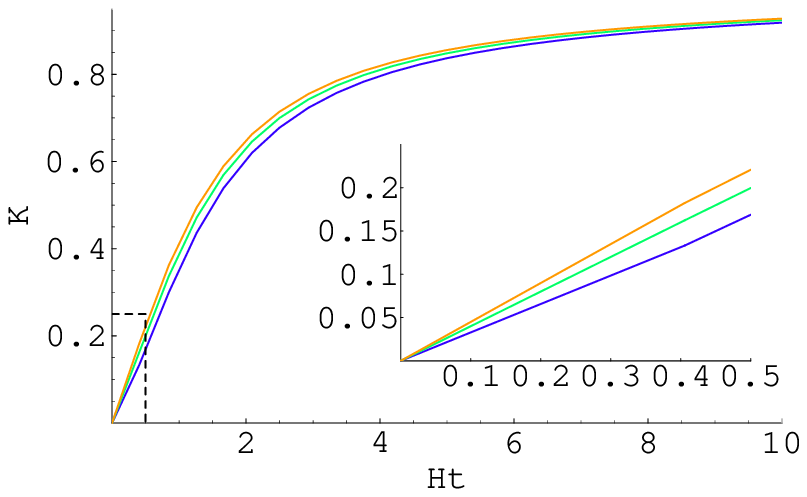}}
\caption{Left panel: plot of the variance $\<\df^2(t)\>$ as a function of 
$\Delta N=H(t-t_{60})$, in the non-Markovian (thin continuous lines) and 
Markovian (thick dashed line) case.
Right panel: plot vs. $\Delta N$ of the ratio $\frac{\<\df^2(t)\>|_{NM}}
{\<\df^2(t)\>|_M}$ between the non-Markovian ($NM$) and Markovian ($M$) case. 
The inset contains a magnification of the region enclosed in the dashed box,
showing in detail the behaviour for small $\Delta N$ (\emph{i.e.} 
close to $\ts$). In all plots, different continuous curves represent values of 
$H(t_{60}-t_{in})$ corresponding to 1.5, 2.5 and 5 (from bottom to top).
The conditional variance is at any time after $\ts$ an increasing function of
the number of e-folds before $\ts$, converging to the value of the 
unconditional variance \eqref{varianza}}
 \label{varianzafig}
\end{figure*}

\vspace{-.3cm}

\begin{figure*}
  \centering
  \psfrag{Ht}{$\Delta N$}
  \psfrag{K}{$K$}
  \subfigure{\kern -.9em
  \includegraphics[width=.45\textwidth,height=5.4cm]{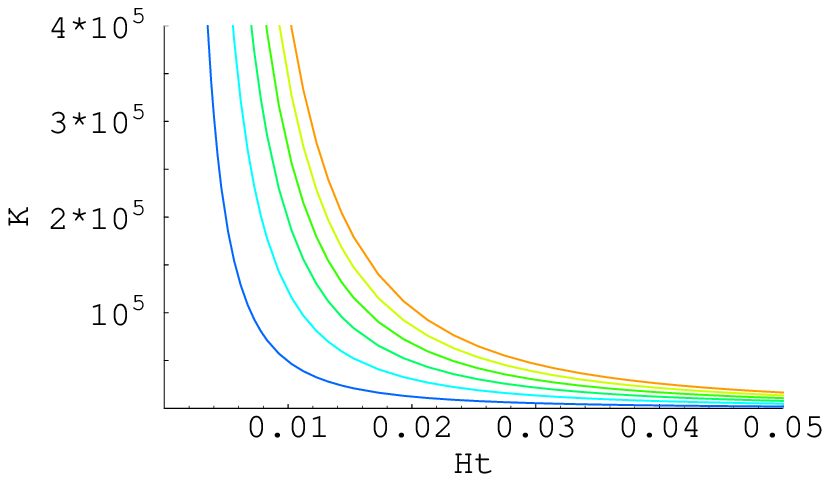} }
  \hspace{.07\textwidth}
  \kern -2em
  \subfigure{\raisebox{.2cm}{
  \includegraphics[width=.41\textwidth,height=5.1cm]{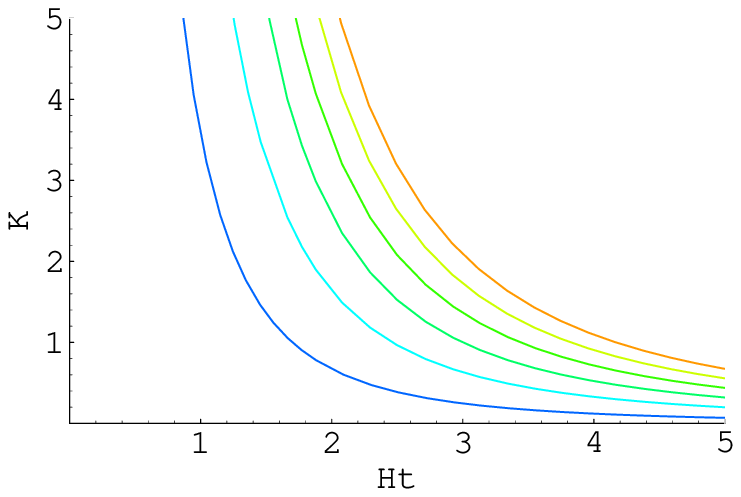}}}
        \\
  \vspace{-.3cm}
  \hspace*{.2cm}
  \subfigure{\psfrag{K}{\kern -3em\raisebox{0cm}{
       $((2\pi)^4/\mu H^2)\,\<\df^3\>$}}
   \includegraphics[width=.4\textwidth,height=.30\textwidth]{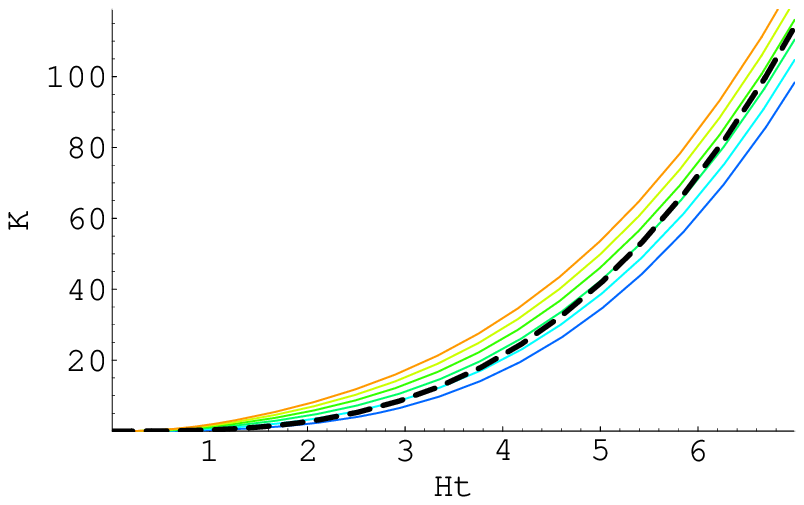}}
  \hspace{.06\textwidth}
  \subfigure{\psfrag{K}{\kern -3em\raisebox{0cm}{ 
       $\<\df^3\>|_{NM}\Big/\<\df^3\>|_M$}}
   \includegraphics[width=.41\textwidth,height=.30\textwidth]{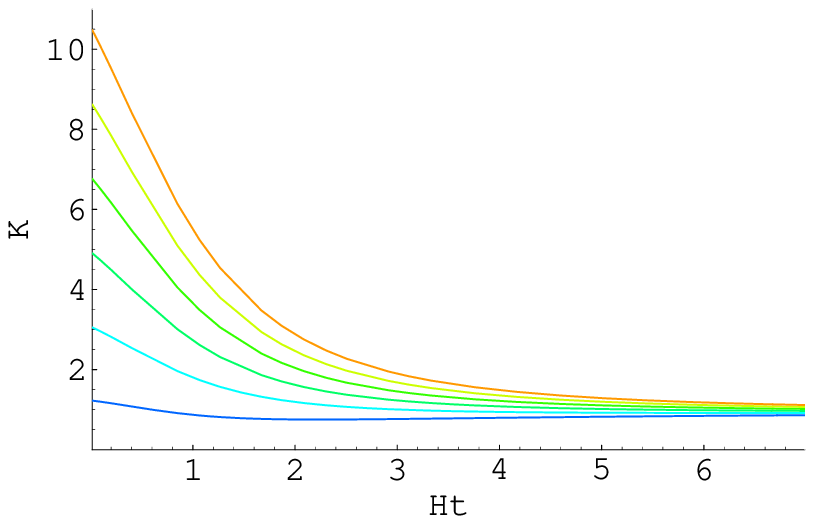}}
\caption{Top panels: plot of the $K$ coefficient for the Gaussian window 
vs. $\Delta N$, showing memory effects in the third moment $\<\df^3(t)\>$, 
over two different time scales.
Bottom panels: dependence on $\Delta N$ of the third moment of $\df$ (left 
panel), in the non-Markovian (thin continuous lines) and Markovian 
(thick dashed line) case; 
behaviour of the ratio $\frac{\<\df^3(t)\>|_{NM}}{\<\df^3(t)\>|_{M}}$ 
(right panel) between the non-Markovian ($NM$) and Markovian ($M$) case. 
Different curves correspond to values of 
$H(t_{60}-t_{in})$ from 5 (bottom line) to 30 (top line).}
\label{memoria}
\end{figure*}

We thus get for the $y$ parameter
\begin{equation}
  y=\frac{C_{12}^2}{C_{11}C_{22}}\sim\frac{1}{2}\frac{1}{H(\ts-\tin)},
\end{equation}
and the variance becomes
\begin{equation}
  \<\df^2(t)\>|_{NM}=\frac{H^4}{8\pi^2}\!\bigg(\!1-\frac{1}{2H(\ts-\tin)}\bigg)
  (t-\ts)^2.
\end{equation}
This expression shows two distinct effects. The first is the fact that 
the variance is an increasing function of the total amount of
e-foldings of inflation before the constraint, and since $y$ tends to 
vanish (though not very fast) for $\ts-\tin\gg H^{-1}$ its dependence 
saturates to the value of the unconditional variance \eqref{varianza}. This is 
a physical consequence of the fact that, even though highly oscillating 
noise configurations are statistically suppressed due to the 
non-vanishing correlation time, if there is much time after 
the beginning of inflation the noise about $\ts$ is almost 
uncorrelated with its initial value, while if the starting time 
is closer to $\ts$ there cannot be very large fluctuations of 
the noise $\xi$, and also the variance of $\df$ gets smaller.
A second effect is the quadratic time dependence soon after the constraint, 
versus the linear dependence of the standard Markovian case \eqref{2pstandard}.
This behaviour, already seen in Eq. \eqref{varianza}, does not have 
anything to do with memory effects, but is merely due to the local shape 
of the noise correlation function, which is never divergent. 
Actually, we still have that 
\begin{equation}
  \frac{\<\df^2(t)\>|_{NM}}{\<\df^2(t)\>|_M}\sim\frac{\Delta N}{2},
\end{equation}
where we have defined $\Delta N= H(t-\ts)=60-N$, with $N$  the number
of e-folds till the end of inflation (see the inset in Figure 
\ref{varianzafig}). This result shows that the power-spectrum
of perturbations on large scales is naturally bluer than the standard 
one. This basic conclusion is in qualitative agreement with the
recent {\it WMAP} results which show a suppression of the lower multipoles 
in the Cosmic Microwave Background (CMB) anisotropies.

\begin{figure*}
  \centering
  \psfrag{Ht}{$\Delta N$}
  \subfigure{\psfrag{K}{$ (H/\mu)\,R_{3/2}$} \kern -.5em
    \includegraphics[width=.4\textwidth,height=.3\textwidth]{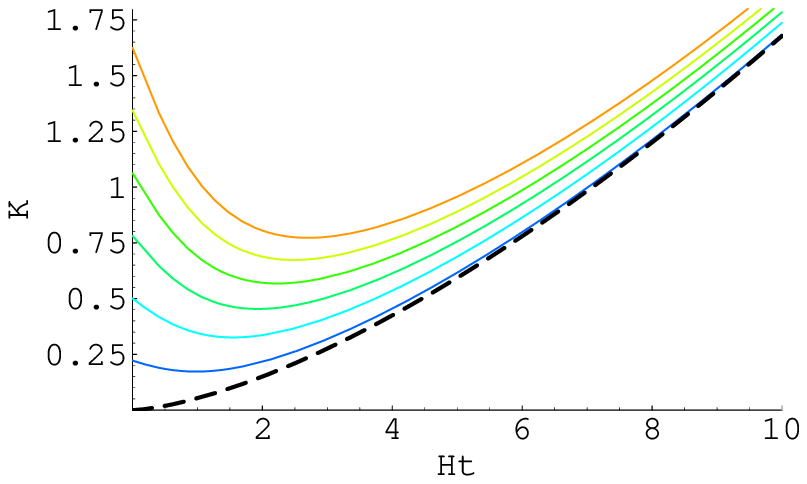}}
  \hspace{.015\textwidth}
  \subfigure{\psfrag{K}{$(H^2/\mu)\,R_{2}$} \raisebox{.1cm}{
    \includegraphics[width=.4\textwidth,height=.3\textwidth]{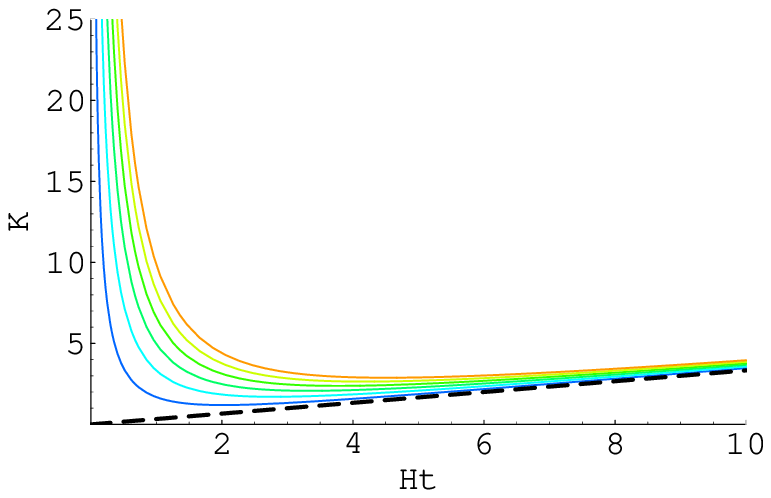}}}
        \\
  \subfigure{\psfrag{K}{\kern -1em $R_{3/2}^{(NM)}\big/R_{3/2}^{(M)}$}
   \includegraphics[width=.4\textwidth,height=.3\textwidth]{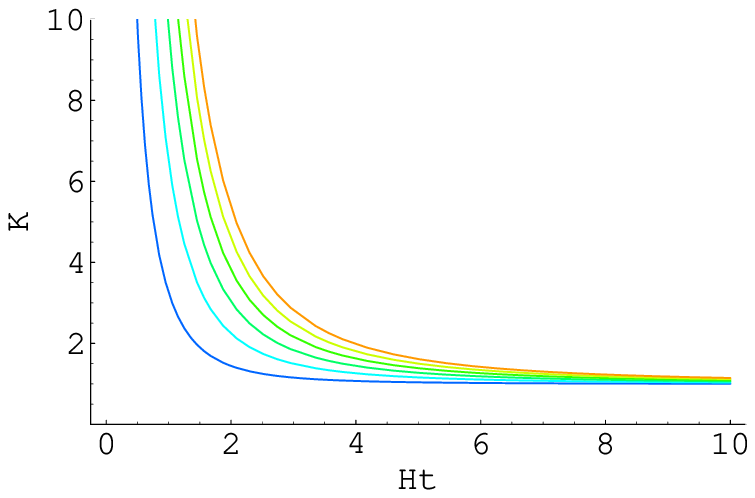}}
  \hspace{.01\textwidth}
  \subfigure{\psfrag{K}{\kern -1em $R_{2}^{(NM)}\big/R_{2}^{(M)}$}
   \includegraphics[width=.4\textwidth,height=.3\textwidth]{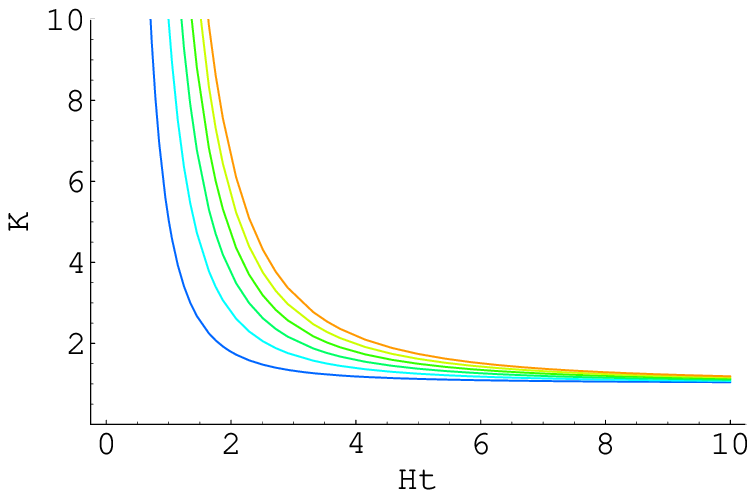}}
  \caption{Top: Behaviour of $R_{3/2}$ (left panel) and $R_{2}$ (right panel) 
as a function of $\Delta N$, for the non-Markovian (thin continuous lines) and 
Markovian (thick dashed line) case. Bottom: behaviour of the ratios 
$R_{3/2}^{(NM)}\big/ R_{3/2}^{(M)}$ (left panel) and 
$R_{2}^{(NM)}\big/R_{2}^{(M)}$ (right panel) of the two cases. In each plot, 
different continuous curves represent values of $H(t_{60}-t_{in})$ 
varying from 5 (bottom line) to 30 (top line).}
\label{nongauss}
\end{figure*}

While memory effects do not clearly show up in the variance, for the third 
moment we obtain, since $K$ is dominated by the $D_{222}$ term (see the 
explicit computation of all the $D_{ijk}$ coefficients for this choice of 
the window in Appendix \ref{coefficienti}),
\begin{align}
  \<\df^3(t)\>|_{NM} 
  &\simeq -\frac{\mu}{H}\frac{3C_{11}C_{12}}{2C_{22}^2}D_{222} \nonumber \\
  &\sim-\frac{\mu H^6(\ts-\tin)}{8\left(2\pi\right)^4}(t-\ts)^3,
\end{align}
and therefore (recalling \eqref{trepuntistandard})
\begin{equation}
  \frac{\<\df^3(t)\>|_{NM}}{\<\df^3(t)\>|_M} \sim \frac{3}{8}H(\ts-\tin),
\end{equation}
showing that for wavelengths slightly smaller than the present 
horizon, the third moment can be considerably 
larger than it is usually assumed, depending on the previous duration of 
inflation. However, at later times (for $\Delta N\gg 1$) both the
variance and the third moment converge to the standard Markovian behaviour. 
The precise value of all these quantities for generic times is shown
in Figure \ref{varianzafig} and Figure \ref{memoria}.
}

The amount of non-Gaussianity can be evaluated from the ratio of the 
three-point function and the variance to the appropriate power.
It is useful to consider both the quantities 
$R_{3/2}\equiv\frac{\<\df^3(t)\>}{\<\df^2(t)\>^{3/2}}$ and 
$R_{2}\equiv\frac{\<\df^3(t)\>}{\<\df^2(t)\>^2}$ (the second one is 
proportional to the non-Gaussianity strength parameter $f_{NL}$, 
see {\it e.g.} Refs. \cite{acqua,malda}) as a function of 
$H(t-t_{60})$ and $H(t_{60}-t_{in})$. 
In this case too, as we see in Figure \ref{nongauss}, the effect  
increases as the time $t$ gets closer to $t_{60}$ and farer from 
$t_{in}$. Analytically, in this limit we have
\begin{gather}
  R_{3/2}^{(NM)}\sim\frac{\mu}{H}\frac{H(\ts-\tin)}{4\sqrt{2}\pi},\\
  R_{2}^{(NM)}\sim \frac{\mu}{H^2}\frac{H(\ts-\tin)}{2\Delta N} \;,
\intertext{(where the superscript $NM$ stands for non-Markovian) while 
for the white-noise case}
  R_{3/2}^{(M)}=\frac{\mu}{H}\frac{\left(\Delta N\right)^{3/2}}{6\pi},\\
  R_{2}^{(M)}=\frac{\mu}{H^2}\frac{\Delta N}{3} 
\end{gather}
(where the superscript $M$ stands for Markovian).
In the Markovian case both quantities vanish, but in non-Markovian ones they 
are finite or even divergent for $\Delta N\rightarrow0$, because of the 
quadratic (instead of linear) time dependence of the variance, and are 
growing functions of $H(t_{60}-t_{in})$, due both to the increase of the third
moment and the decrease of the variance with the total number of e-folds
before $\ts$. 
Non-Gaussianity can then be larger by {\it orders of magnitude} in our
scheme.


\section{Discussion and conclusions}

\label{conclusioni}

In this paper we used the stochastic inflation approach to address an 
important theoretical issue, namely that of estimating the possible 
influence of super-horizon perturbation modes on the statistics of 
cosmological perturbations on observable sub-horizon scales. 
To this aim we first modified the standard scheme in order to allow for 
the cross-talk of perturbations on super and sub-horizon scales. 
Such an effect is indeed totally absent in the traditional stochastic 
inflation dynamics; this is an artifact of the sharp k-space filter adopted 
in the coarse-graining procedure, which is easily removed by adopting a 
smoother filter function, leading to a colored - rather than white - noise 
source in the Langevin-like equation which governs the evolution of the 
coarse-grained inflaton field. This modification implies that the evolution of 
the coarse-grained inflaton field behaves as a non-Markovian stochastic 
process, in contrast to the standard case. 
Perturbations relevant to our smooth local patch of the Universe are then 
consistently defined by constraining the inflaton field to be  
homogeneous at a conventional time $t_{60}$ ({\it i.e.} about 60 e-folds 
before inflation ends), corresponding to the horizon crossing of a scale 
slightly larger than the present Hubble radius. As a consequence of its 
non-Markovianity, the coarse-grained inflaton field preserves some memory of 
its dynamics prior to the constraint, which means that sub-horizon-scale 
perturbations have some knowledge of the state of the Universe on 
super-horizon scales.  

Endowed with this {\it extended} stochastic inflation scheme we are finally 
able to calculate the conditional second- and third-order moments of inflaton 
perturbations on observable scales. We perform our calculations in the case 
of a simple model where a scalar field with a small cubic 
self-interaction term evolves in a fixed de Sitter 
background. Our most important findings are: 
\begin{itemize}
\item The variance of inflaton fluctuations grows quadratically with time 
around $t_{60}$, and is therefore smaller than in the standard Markovian 
case, which is linearly dependent on time; this is equivalent to 
say that the power-spectrum of density perturbations gets bluer on 
very large scales, without invoking any ad-hoc new physical input.
This generic feature looks very intriguing, since 
the CMB anisotropy power on the largest angular scales observed by 
{\it WMAP} \cite{spergel} appears to be lower than the one predicted by 
the standard model of cosmology with almost scale-invariant primordial 
perturbations arising from a period of inflation.

\item The skewness of inflaton fluctuations is larger than the standard case 
at times around $t_{60}$, which is equivalent to an enhancement of the 
non-Gaussianity level on large scales, for a given value of the self-coupling 
strength. The possible presence of large-scale non-Gaussianity is 
an important prediction to be confronted with current observational limits 
\cite{komatsu}. Let us also mention that some recent analyses have
reported evidence for a positive detection of non-Gaussianity in the 
{\it WMAP} 1-year data 
\cite{ngdetect1,ngdetect2,ngdetect3,ngdetect4,ngdetect5}.      
\end{itemize}

In spite of the simplicity of the considered model, we can  
conclude that the cross-talk between super- and sub-horizon-scale 
perturbations is an important effect which would deserve an accurate 
treatment. The scheme presented here should be considered only as a first step
in this direction, which needs to be largely improved and extended in 
various ways. 
First of all, one should consistently include metric perturbations in the 
inflaton dynamics, going beyond our fixed de Sitter background treatment.
Second, our choice of a Gaussian filter can be somewhat arbitrary, 
and it would be important to understand how much the results depend on 
its choice. The conclusions reached in Ref. \cite{Winitzki}, according to 
which there is a wide class of filters whose noise correlators 
show the same asymptotic behaviour $\<\xi(t)\xi(t')\>\propto e^{-2H|t-t'|}$
found in this paper (see Eq. \eqref{limnoisecorr}), 
are however quite encouraging in this respect. 
Actually, since the window function is only a technical device, 
we would expect our results to be independent of its choice, at least
qualitatively. The white-noise exception is not significant, 
being merely a consequence of the ``bad'' choice of smoothing in 
configuration space. 

The existence of memory effects means that the detailed  
dynamics of inflation plays a role in the specific form assumed by 
observable quantities. For instance, in our model one finds a residual 
dependence on the time when inflation started, $t_{in}$. 
Nonetheless, it is quite likely that this dependence saturates to a universal 
value determined by the general asymptotic behaviour, if the overall number 
of inflation e-folds is very large, as it is usually the case. 
More difficult is to avoid a dependence on the precise time 
$t_{60}$ at which we put the homogeneity constraint, and on the specific 
form of the constraint, which we introduced here through a delta function
in  field configuration space 
(see, however, Ref. \cite{lesarcs} for a discussion of alternative 
approaches to the constraint). 
These are important issues which will certainly deserve further 
investigation.


\appendix


\onecolumngrid

\vspace{1cm}

\section{Stochastic noise and Langevin equation}

\label{appA}

The effective equation of motion for the {\it in-in} expectation value of the 
super-horizon field $\varphi_<$ is obtained via a path-integral over the 
sub-horizon field $\varphi_>$:
\begin{equation}
  \kern -1em e^{i\Gamma[\ppmmin]} =  e^{i[S^+_< - S^-_<]}\! 
  \int\!\!\mathcal{D}\ppmmag e^{i\!\int\! dx \left[\frac{1}{2}\ppmag(x) 
  \Lambda(x)\ppmag(x) - \frac{1}{2}\pmmag(x) \Lambda(x)\pmmag(x) + 
  \ppmin(x)\Lambda(x)\ppmag(x) - \pmmin(x)\Lambda(x)\pmmag(x)\right]},
\end{equation}
where $\Lambda$ is the integration kernel of the action for a free scalar 
field, given by
\begin{equation}
  \Lambda(x)=-a^3_t\left(\de_t^2+3H\de_t-\frac{\nabla^2}{a^2}+m^2\right).
\end{equation}

After some manipulations \cite{morikawa}, setting $\ppmmin=\varphi^\pm$, 
$\varpd=\ppmin-\pmmin$ and $\varpc=\frac{1}{2}(\ppmin+\pmmin)$, the effective 
action reads
\begin{multline}
\label{azioneeff}
  \Gamma[\varphi^\pm] = S[\varphi^+]-S[\varphi^-]
  +\frac{i}{2}\!\int\!\!d^4xd^4x'\varpd(x)\mathrm{Re}[\Pi(x,x')]\varpd(x')
  -2\!\int\!\!d^4xd^4x'\theta(t-t')\varpd(x)\mathrm{Im}[\Pi(x,x')]\varpc(x'),
\end{multline}
where 
\begin{gather}
  \Pi(x,x')\equiv\!\int\!\!\dk a_t^3\hat{P}_t\phi_\mathbf{k}(x)
    a_{t'}^3\hat{P}_{t'}\phi^*_\mathbf{k}(x')\quad,\quad
  \hat{P}_t=\ddot W +3H\dot W +2\dot W\partial_t
\end{gather}
and the normal modes $\phi_\mathbf{k}$ of the field are given by 
(\ref{mnormali}).

Thanks to the useful relations $\:a_t^3\hat{P}_t=\partial_t(a_t^3 \dot 
W_t(k))+2a_t^3\dot W_t(k)\partial_t\:$ and
\begin{equation}
\label{qnu}
  \dot\phi_\mathbf{k}=-H\left(\frac{3}{2}-\nu+\frac{k}{aH}
  \frac{H_{\nu-1}^{(1)}\left(\frac{k}{aH}\right)}{H_\nu^{(1)}\left(
  \frac{k}{aH}\right)}\right)\phi_\mathbf{k}\equiv q_\nu(k\eta)\phi_\mathbf{k},
\end{equation}
after integrating by parts, the imaginary term of the effective action 
(\ref{azioneeff}) reads
\begin{equation}
\label{immag}
  \frac{i}{2}\!\int\!\!\dx\dx'a_t^3a_{t'}^3\mathrm{Re}\!\int\!\!\dk 
[\varpd(x)q_\nu(k\eta)-\pddot(x)] 
  \dot W_t(k)\Phi_\mathbf{k}(x)\dot W_{t'}(k)\Phi^*_\mathbf{k}(x')
  [\varpd(x')q_\nu^*(k\eta')-\pddot(x')].
\end{equation}

Assuming $W_t(k)=W(k\eta)$, then $\dot 
W_t=(k/a_t)W'(k\eta)$, where the prime denotes differentiation with respect 
to the argument of $W$, and with this substitution we obtain
\begin{equation}
  \dot W_t(k)\Phi_\mathbf{k}(x)\dot W_{t'}(k)\Phi^*_\mathbf{k}(x')=
  \frac{k^2e^{i\mathbf{k\cdot(x-x')}}}{32\pi^2H}
  W'(k\eta)W'(k\eta')
  \frac{H_\nu^{(1)}(\frac{k}{a_tH})H_\nu^{(1)*}(\frac{k}{a_{t'}H})}
  {(a_ta_{t'})^{5/2}};
\end{equation}
we now set $\psi^\Delta(x)=(\frac{3}{2}-\nu)\pd_<(x)+\dot\phi^\Delta_<(x)/H$, 
and from the explicit form (\ref{qnu}) of $q_\nu$ (after evaluating the 
integral over angles) we can write the integral over $\dk$ in 
a matrix form as
\begin{gather}
  [\psi^\Delta(x),\varpd(x)]\frac{H}{8\pi}\!\int\!\!dk\:k^4 \frac{\sin kr}
  {kr}\frac{W'(k\eta)W'(k\eta')}{(a_ta_{t'})^{5/2}}\mathrm{Re}
  [\mathbf{M}_\nu(k\eta,k\eta')]
  \left[\!\!
  \begin{array}{c}     
    \psi^\Delta(x')\\ \varpd(x')
  \end{array}\!\!
  \right],
\intertext{with the matrix $M^{i,j}_\nu(k\eta,k\eta')$ given by 
(\ref{Mnu}). This matrix is Hermitian under the simultaneous exchange of 
discrete and continuous indices $i,t\rightarrow j,t'$, and 
Re$[\mathbf{M}_\nu]$ is therefore symmetric.
Eq. (\ref{immag}) can thus be rewritten in the bilinear and symmetric 
form}
  \frac{i}{2}\!\int\!\!\dx\dx'a^3_ta^3_{t'}[\psi^\Delta(x),\varpd(x)]
   \mathbf{A}(x,x')\left[\!\!
  \begin{array}{c}     
    \psi^\Delta(x')\\ \varpd(x')
  \end{array} \!\!\right],
\intertext{where $\mathbf{A}(x,x')$ is the correlation matrix 
(\ref{intdk}); introducing two classical random fields $\xi_1$ and $\xi_2$ 
with the statistical weight (\ref{pesogaussiano}), for the imaginary part 
of the effective action (\ref{azioneeff}) we finally get}
  e^{-\frac{1}{2}\!\int\!\!d^4xd^4x'\pdmin(x)\mathrm{Re}[\Pi(x,x')]
  \pdmin(x')}=\int\!\!\mathcal{D}\xi_1 
\mathcal{D}\xi_2 P[\xi_1,\xi_2]e^{i\int\!\dx a^3_t 
H\left[\psi^\Delta(x)\xi_1(x)+\varpd(x)\xi_2(x)\right]},
\end{gather}
from which we immediately obtain (\ref{gamma}).


\section{dissipation}
\label{dissipaz}

In the Langevin theory of Brownian motion, the stochastic force arising
as a consequence of the collisions with the particles of the thermal bath is 
usually split in two parts, one rapidly varying and stochastically 
distributed and another one proportional to the particle velocity, which plays
the role of a friction term. The latter contribution describes how fast 
the system reaches thermal equilibrium from an out-of-equilibrium 
configuration, making the mean velocity vanish exponentially and 
driving the mean kinetic energy to the equipartition value. 
The proportionality coefficient of this friction term (and therefore the 
characteristic relaxation time needed for the system to reach equilibrium and 
for the friction to become negligible) can be derived integrating over time the
correlation function of the rapidly varying part of the stochastic force.
This relation between the macroscopic out-of-equilibrium behaviour and the
microscopic equilibrium distribution is known as the 
{\it fluctuation-dissipation} theorem.

We then expect in our situation too a dissipation term of the type $\alpha H
\dot\varphi$ to show up in the equation of motion, with the coefficient 
$\alpha$ related in some way to the correlation functions of the noise $\xi$. 
We also expect this effect to be negligible after a certain time.

In Sec. \ref{derivazeqmoto} we disregarded the real term that appears in the 
action with the path-integral over the sub-horizon degrees of freedom (the 
$\im[\Pi]$ term in (\ref{azioneeff})). Had we kept this term, after integrating
by parts with respect to $t'$ we would have obtained in the equation of motion 
the two extra contributions
\begin{equation}
  \label{eqdiss}
  \int \dx'a_{t'}^3\vartheta(t-t')\zeta(x,x')\dot\varphi(x')
  +\int\dx'a_{t'}^3\delta m^2(x,x')\varphi(x'),
\end{equation}
corresponding to non-local effects of dissipation and mass renormalization 
respectively, with the integration kernels given by
\begin{gather}
  \zeta(x,x')=2\,\im\!\int\!\dk\:\hat P_t\phi_\mathbf{k}(x)\:\dot W_{t'}
  \phi^*_\mathbf{k}(x'),\\
  \delta m^2(x,x') = -4 \delta(t-t')\,\im\!\int\!\dk\:\dot W_t
  \dot\phi_\mathbf{k}(x)\dot W_{t'}\phi^*_\mathbf{k}(x')-2\,\vartheta(t-t')\im
  \!\int\!\dk\:\hat P_t\phi_\mathbf{k}(x)\:\dot W_{t'}\dot\phi^*_\mathbf{k}(x')
  ,
\end{gather}
where in $\delta m^2$ a second term proportional to $\delta(t-t')
\phi_\mathbf{k}(x)\phi^*_\mathbf{k}(x')$ was dropped out since it gives a 
purely real result.

We will concentrate on the first term of \eqref{eqdiss}, showing that it is 
negligible compared to the usual friction term $3H\dot\varphi_\mathbf{k}$ 
related to the Hubble expansion. Since it has no space dependence but 
$e^{i\mathbf{k}\cdot(\mathbf{x-x'})}$, going to Fourier space we have
\begin{equation}
\int_{-\infty}^t \!\!\! dt'a_{t'}^3\frac{H^5}{k^3}\frac
{k^4\eta^2{\eta'}^2}{\sigma^4}\im\bigg[e^{-\frac{(k\eta)^2}
{2\sigma^2}-ik\eta}
 e^{-\frac{(k\eta')^2}{2\sigma^2}+ik\eta'}
\bigg((1+ik\eta)\bigg(1+\frac{k^2\eta^2}{\sigma^2}\bigg)-2k^2\eta^2\bigg)
(1-ik\eta')\bigg]\dot\varphi_\mathbf{k}(t').
\end{equation}
The field $\varphi$ basically contains super-horizon modes, which are slowly 
varying with respect to the characteristic correlation time of the sub-horizon 
fluctuations. We can then assume that $\dot\varphi_\mathbf{k}(t')$ 
is constant where the rest of the integrand is significantly different from 
zero, and take it out from the integration. We thus obtain a dissipation term 
$\alpha_\mathbf{k}H\dot\varphi_\mathbf{k}(t)$ for the $k$ mode, with a time 
dependent friction coefficient $\alpha_\mathbf{k}=\alpha(k\eta,\sigma)$. 
Changing variables, it becomes 
\begin{equation}
  \alpha(k\eta,\sigma) = 
  \frac{k^2\eta^2}{\sigma^4}\im\bigg[e^{-\frac{(k\eta)^2}{2\sigma^2}-ik\eta}
  \bigg((1+ik\eta)\bigg(1+\frac{k^2\eta^2}{\sigma^2}\bigg)-2k^2\eta^2\bigg)
  \int_{-\infty}^{k\eta} \!\!\! dx e^{-\frac{x^2}{2\sigma^2}+ix}
  \frac{1-ix}{x^2}\bigg];
\end{equation}
this fluctuation-induced dissipation effect can then be neglected when 
$|\alpha(k\eta,\sigma)|\ll3$. For small values of $k|\eta|$ (sufficiently after
the horizon crossing of each mode), this inequality reduces to 
\begin{equation}
  \frac{k^2\eta^2}{\sigma^2}\bigg(\frac{1}{3}-\frac{\sigma^2}{3\cdot5}
  +\frac{\sigma^4}{3\cdot5\cdot7}+\cdots\bigg) \ll 3,
\end{equation}
and is always satisfied for $k|\eta|\lesssim\sigma\leq1$.

\begin{figure}
  \centering
      \psfrag{a}{\raisebox{-5mm}{$\kern -1em \alpha$}}
    \psfrag{k}{\raisebox{-3mm}{$\kern -1em k|\eta|$}}
    \includegraphics[width=.4\textwidth]{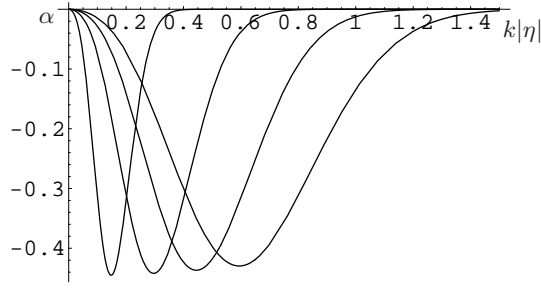} 
    \caption{Friction coefficient $\alpha(k\eta,\sigma)$ governing the 
fluctuation-induced dissipation effect for the $k$ mode, for different values 
of $\sigma$ (from left to right, $\sigma=.125,\;.25,\;.375,\;.5$). The sign of 
the coefficient is always negative (meaning energy gain), and all functions are
peaked roughly around the cutoff scale of the mode, for $k|\eta|\sim\sigma$ }
\label{dissipfig}
\end{figure}

The complete behaviour of the friction coeffiecient $\alpha$ is shown in 
Figure \ref{dissipfig}, where we see that for each mode the dissipative 
behaviour reaches its maximum for $k|\eta|\sim\sigma$, when the mode 
crosses the effective horizon $(\sigma H)^{-1}$; the spatial non-locality 
of the dissipation term in \eqref{eqdiss} is a consequence of this 
effectiveness of the friction coefficient only about the cutoff scale 
for each mode 
\cite{morikawa}. We also note that the sign is always negative: this 
``anti-dissipation'' means for $\varphi$ a global energy gain, instead of a 
loss, and is due to the continuous income of modes through the cutoff into the
super-horizon field. Anyway, this effect can be neglected on a first approach, 
since $\alpha/3$ is always much less than unity.


\section{Noise correlation matrix}
\label{noisecorr}

The noise correlation matrix $A_{ij}(x,x')$ can be calculated in the 
Gaussian window case (\ref{finestra}). Since the relation
\begin{equation}
  \sin(kr)\mathrm{Re}[\mathbf{M}_\nu]=
  \frac{1}{2}\mathrm{Im}[\mathbf{M}_\nu(e^{ikr}-e^{-ikr})]
\end{equation}
holds, Eq. (\ref{intdk}) then becomes (setting $\delta=\eta'-\eta$ and 
$R\equiv\sqrt{\eta^2+{\eta'}^2}$)
\begin{align}
  \mathbf{A}(x,x')&=\frac{H^6}{8\pi^2 
r}(\eta\eta')^3\mathrm{Im}\!\int_0^\infty\!\!dk\:k^5 
\frac{e^{-\frac{\eta^2+{\eta'}^2}{2\sigma^2}k^2}}{\sigma^4}\!
  \mat{cc}{\frac{(k\eta-i)(k\eta'+i)}{\eta\eta'k^3} \!&\! 
\eta'\frac{1+ik\eta}{k\eta} \nonumber \\
           \eta\frac{1-ik\eta'}{k\eta'} \!&\! 
  \eta\eta'k}\!(e^{i(\delta+r)k}-e^{i(\delta-r)k})\\
  &=\frac{H^6(\eta\eta')^3}{8\pi^2 r\sigma^4}\Big[\phantom{2} 
  \boldsymbol{\de} \phantom{2}\Big] 
  \frac{\de^2}{\de\delta^2}\mathrm{Im}\!\int_0^\infty\!\!dk 
  e^{-\frac{\eta^2+{\eta'}^2}{2\sigma^2}k^2}\!(e^{i(\delta+r)k}-
  e^{i(\delta-r)k})\\
  &=\frac{\sqrt{2}H^6(\eta\eta')^3}{8\pi^2 R r \sigma^3} \Big[\phantom{2} 
\boldsymbol{\de} \phantom{2}\Big] \frac{\de^2}{\de\delta^2}
  \!\!\left[e^{-\frac{(\delta+r)^2\sigma^2}{2 
R^2}}\!\!\int_0^{(\delta+r)\sigma/\sqrt{2}R}\kern-1.5em dy 
e^{y^2}-e^{-\frac{(\delta-r)^2\sigma^2}{2R^2}}\!
  \int_0^{(\delta-r)\sigma/\sqrt{2}R}\kern-1.5em dy e^{y^2}\right], \nonumber
\end{align}
where we have defined the derivation matrix
\begin{equation}
  \Big[\phantom{2} \boldsymbol{\de} \phantom{2}\Big]=
    \mat{cc}{\displaystyle 
\dd{2}+\frac{1}{\eta\eta'}\left(\delta\dd{}-1\right) & \displaystyle 
\frac{\eta'}{\eta}\left(1+\eta\dd{}\right)\dd{2} \vspace{5pt}\\
    \displaystyle\frac{\eta}{\eta'}\left(1-\eta'\dd{}\right)\dd{2} & 
\displaystyle-\eta\eta'\dd{4}}.
\end{equation}

For $r\rightarrow0$ we have
\allowdisplaybreaks{
\begin{align}
  \mathbf{A}(t,t')_{r=0} &= -\frac{\sqrt{2}H^6(\eta\eta')^3}{4\pi^2 
R^2\sigma^2}\Big[\phantom{2} \boldsymbol{\de} \phantom{2}\Big] 
\frac{\de^2}{\de\delta^2}
  \!\left[\frac{\delta\sigma}{R} e^{-\frac{\delta^2\sigma^2}{2R^2}}
  \!\!\int_0^{\delta\sigma/\sqrt{2}R} \kern-1.5em dy e^{y^2}\right] \nonumber\\
  &= -\frac{H^6}{4\pi^2}\frac{(\eta\eta')^3}{R^4} \Big[\phantom{2} 
\boldsymbol{\de} \phantom{2}\Big]
  2\sum^\infty_{k=0}\frac{(-1)^k(k+1)}{(2k-1)!!}\left(
    \frac{\delta\sigma}{R}\right)^{2k}\\
  &=\frac{H^6}{4\pi^2}\left(\!\frac{2\eta\eta'}{\eta^2+{\eta'}^2}
  \!\right)^{\!\!2}\sum^\infty_{k=0}\mathbf{\mathcal{A}}^{(k)}(t-t')
  \frac{(-1)^k(k+1)(k+2)}{(2k-1)!!}\left(
  \frac{(\eta'-\eta)^2}{\eta^2+{\eta'}^2}\sigma^2\right)^{k}, \nonumber
\end{align}
($\mathbf{\mathcal{A}}^{(k)}(t-t')$ is given in (\ref{matrice})), and 
since in de Sitter space $\eta=-e^{-Ht}/H$ we get (\ref{serie}).
}


\section{Conditional probability}

\label{probabilita}

As a shorthand notation to skip writing all the integration symbols, we 
define the inner product of two functions $f(x)$ and $g(x)$ as
\begin{gather}
  f^Tg=\int_{-\infty}^{+\infty}\dx f(x)g(x);
\intertext{in this notation the equations of motion (\ref{eqpert1}) and 
(\ref{eqpert2}) become}
  \df[\xi](t) = \df_{60} - \frac{\mu}{6H}\df_{60}^2(t-\ts) + J^T_1\xi - 
\frac{\mu}{6H}\xi^T\B{1}\xi \;\;,\;\; \df[\xi](t_{60}) = J^T_2\xi - 
\frac{\mu}{6H} \xi^T\B{2}\xi 
\intertext{with (the function $\vartheta(t<t'<t'')$ is defined to be 1 if 
the inequality is true, 0 elsewhere)}
   J_1(x')=\vartheta(t_{60}<t'<t)\delta(\mathbf{x}-\mathbf{x}') 
\quad,\quad 
J_2(x')=\vartheta(t_{in}<t'<t_{60})\delta(\mathbf{x}-\mathbf{x}') \\
  \B{1} (x',x'')= \delta(\mathbf{x}-\mathbf{x}')\delta(\mathbf{x}-
   \mathbf{x}'')\int_{t_{60}}^{t}\!\!\!d\tilde t 
   \vartheta(t_{60}<t'<\tilde t)\vartheta(t_{in}<t''<\tilde t)\\
  \B{2}(t',t'')= \delta(\mathbf{x}-\mathbf{x}')\delta(\mathbf{x}-
    \mathbf{x}'')\int_{t_{in}}^{t_{60}}\!\!\!d\tilde t
   \vartheta(t_{in}<t'<\tilde t)\vartheta(t_{in}<t''<\tilde t)
\end{gather}

Since we want a probability distribution for connected correlation 
functions, as discussed in (\ref{iteraz3}) we subtract the mean of 
the perturbative solution, which in this notation reads 
\begin{equation}
  \<\df[\xi](t)\>=-\frac{\mu}{6H}\tr[\mathbf{A}\B{1}] \quad,\quad 
\<\df[\xi](t_{60})\>=-\frac{\mu}{6H}\tr[\mathbf{A}\B{2}].
\end{equation}

If again we set $\df_1\equiv\df-\df_{60}+\frac{\mu}{6H}\df_{60}^2(t-\ts)$ 
and $\df_2\equiv\df_{60}$, we get for the joint probability 
(\ref{pcongiunta})
\begin{align}
  P(\df,\df_{60}) &= \int\!\!\frac{d\al_1d\al_2}{(2\pi)^2} 
  e^{i\left(\frac{\mu}{6H}\tr[\mathbf{A}\B{i}]-\df_i\right)\al_i}
  \mathcal{N}\!\!\int \!\!\mathcal{D}\xi\:e^{-\frac{1}{2}\xi^T\Ainv
  (\mathbf{1}+i\frac{\mu}{3H}\al_k\mathbf{A}\B{k})\xi+i\al_iJ_i^T\xi}
  \nonumber\\
  &= \int\!\!\frac{d\al_1d\al_2}{(2\pi)^2} 
  e^{i\left(\frac{\mu}{6H}\tr[\mathbf{A}\B{i}]-\df_i\right)\al_i}
  \mathcal{N'}e^{-\frac{1}{2}\al_iJ^T_i(\mathbf{1}+i\frac{\mu}{3H}\al_k
  \mathbf{A}\B{k})^{-1}\mathbf{A}J_j\al_j},
\end{align}
where the new normalization $\mathcal{N'}$ reads
\begin{equation}
  \mathcal{N'}=\frac{\int\!\!\mathcal{D}\xi\:e^{-\frac{1}{2}\xi^T\Ainv
  \left[\mathbf{1}+\frac{i\mu}{3H}\al_k\AB{k}\right]\xi}}{\int \!\!
  \mathcal{D}\xi e^{-\frac{1}{2}\xi^T\Ainv\xi}}=\mathrm{Det}\!
  \left[\mathbf{1}+\frac{i\mu}{3H}\al_k\AB{k}\right]^{-1/2} 
  \equiv e^{-\frac{1}{2}\mathrm{Tr}\:\ln\!\left[\mathbf{1}+
  \frac{i\mu}{3H}\al_k\AB{k}\right]}\;.
\end{equation}

To first order in $\frac{\mu}{6H}$, the normalization $\mathcal{N}'$ and 
the mean value of the field cancel out, and we have:
\begin{equation}
  P(\df,\df_{60}) = \int\!\!\frac{d\al_1d\al_2}{(2\pi)^2} 
\:e^{-i\df_i\al_i}e^{-\frac{1}{2}J^T_i\mathbf{A}J_j\al_i\al_j 
+i\frac{\mu}{6H}J^T_i\ABA{j}J_k\al_i\al_j\al_k};
\end{equation}
normalizing the previous result for the joint probability with the 
probability distribution (\ref{p60}) for $\df_{60}$, that becomes
\begin{equation}
  P(\df_{60}) = \int\!\frac{d\al_2}{2\pi} 
\:e^{-i\df_2\al_2}e^{-\frac{1}{2}J^T_2\mathbf{A}J_2\al_2^2 
+i\frac{\mu}{6H}J^T_2\ABA{2}J_2\al_2^3},
\end{equation}
and uniforming the notation with the one adopted in Section 
\ref{distrprob} (with $J^T_i\mathbf{A}J_j=C_{ij}$ and 
$J^T_i\ABA{j}J_k=D_{ijk}$), we obtain the conditional probability 
distribution (\ref{pcondiz}).


\section{Coefficients $D_{ijk}$ for the Gaussian window}

\label{coefficienti}

We perform here the explicit calculation (in the small $\sigma$ limit) of 
the $D_{ijk}$ coefficients appearing in the conditional probability of 
Section \ref{distrprob}, expanding in the limit $t\rightarrow\ts$ and 
$\ts-\tin\gg H^{-1}$.

\begin{align}
  D_{111} &= -\left(\frac{H}{2\pi}\right)^4\frac{1}{4H}
  \int_{\eta_{60}}^\eta \frac{d\tilde\eta}{\tilde\eta}\left[\ln
  \frac{(\eta^2+\eta_{60}^2)(\tilde\eta^2+\eta_{60}^2)}
  {(\eta^2+\tilde\eta^2)2\eta_{60}^2}\right]^2 
  \simeq\frac{H^8}{12}\frac{1}{(2\pi)^4}(t-\ts)^5 
\end{align}

\begin{align}
  D_{112} = D_{211} &= -\left(\frac{H}{2\pi}\right)^4\frac{1}{4H}
  \int_{\eta_{60}}^\eta\frac{d\tilde\eta}{\tilde\eta}\ln
  \frac{(\eta^2+\eta_{60}^2)(\tilde\eta^2+\eta_{60}^2)}
  {(\eta^2+\tilde\eta^2)2\eta_{60}^2}\ln
  \frac{2\eta_{60}^2(\tilde\eta^2+\eta_{in}^2)}
  {(\eta^2+\tilde\eta^2)(\eta_{60}^2+\eta_{in}^2)} 
  \simeq \frac{H^7}{12}\frac{1}{(2\pi)^4}\,(t-\ts)^4 
\end{align}

\begin{align}
  D_{122} = D_{221} &= -\left(\frac{H}{2\pi}\right)^4\frac{1}{4H}
  \int_{\eta_{in}}^{\eta_{60}}\frac{d\tilde\eta}{\tilde\eta}\ln
  \frac{(\eta_{60}^2+\eta_{in}^2)(\tilde\eta^2+\eta_{in}^2)}
  {(\eta_{60}^2+\tilde\eta^2)2\eta_{in}^2}\ln\frac{(\eta^2+\eta_{in}^2)
  (\eta_{60}^2+\tilde\eta^2)}{(\eta^2+\tilde\eta^2)
  (\eta_{60}^2+\eta_{in}^2)} \nonumber \\
  &\simeq \frac{H^4}{4}\frac{1}{(2\pi)^4}(t-\ts)(\ts-\tin)\ln 2
\end{align}

\begin{align}
  D_{121} &= \frac{H^3}{8}\left(\frac{1}{2\pi}\right)^4\int^1_b
  \frac{dy}{y}\left[\ln\frac{(1+ab)(y+b)}{(y+ab)(1+b)}\right]^2 
  \simeq \frac{H^3}{4}\frac{1}{(2\pi)^4}(t-\ts)^2
  \left(\frac{\ln 2}{2}-1\right)
\end{align}

\begin{align}
  D_{212} &= \frac{H^3}{8}\left(\frac{1}{2\pi}\right)^4\int^1_b
  \frac{dy}{y}\left[\ln\frac{2(1+by)}{(1+y)(1+b)}\right]^2 
  \simeq \frac{H^6}{12}\left(\frac{1}{2\pi}\right)^4 
  (t-\ts)
\end{align}

\begin{align}
  D_{222} &= \frac{H^3}{8}\left(\frac{1}{2\pi}\right)^4\int^1_b
  \frac{dy}{y}\left[\ln\frac{(1+y)(1+b)}{2(y+b)}\right]^2 \nonumber \\
  &\simeq \frac{H^3}{(2\pi)^4}\left[\frac{H^3(\ts-\tin)^3}{3} - 
  \frac{\ln 2}{2}H^2(\ts-\tin)^2 + \!\left(\!\ln^2 2-\frac{\pi^2}{6}\!\right)
  \frac{H(\ts-\tin)}{4}\right]
\end{align}


\end{document}